\documentclass[floatfix,plainnat,tightenlines,prd,nofootinbib,notitlepage,superscriptaddress,10pt,twocolumn]{revtex4-1}

\pdfoutput=1

\usepackage{amsmath}
\usepackage{amssymb}
\usepackage{color}
\usepackage{graphicx}
 \usepackage{notes2bib}
 \usepackage{natbib}
\usepackage{bibentry}

\newcommand{\Christilde}[3]{\ensuremath{ \tilde{\Gamma}^{#1}_{\cdot \, #2 #3}}}

\newcommand{\J}{\ensuremath{\mathrm{J}}}

\newcommand{\h}{\ensuremath{\mathrm{H}}}

\newcommand{\g}{\ensuremath{\mathrm{g}}}
\newcommand{\R}{\ensuremath{\mathcal{R}}}

\newcommand{\LCDM}{\ensuremath{\Lambda \textrm{CDM}}}

\begin{document}
\title{Eternally oscillating zero energy Universe}
\author{Karthik H. Shankar}
\email{Email: kshankar79@gmail.com}
\affiliation{Center for Memory and Brain \\ Boston University}

\begin{abstract}

The question of whether the universe is eternal or if it had a singular moment of creation is deeply intriguing. Although different versions of steady state and oscillatory models of eternal universe have been envisaged, empirical evidence suggests a singular moment of creation at the big bang. 
Here we analyze the oscillatory  solutions for the universe in a modified theory of gravity THED (Torsion Hides Extra-Dimension) and evaluate them by fitting  Type 1a supernovae redshift data.  THED  exactly mimics General Relativity at the kinematical level, while the modifications in its dynamical equations allow the universe to bounce between a minimum size and a maximum size with a zero average energy within each oscillation.  
The optimally fit oscillatory solutions correspond to a universe with (i) a small matter density requiring little to no dark matter, (ii) a significantly negative spatial curvature, (iii) a tiny negative dark energy. 
Alternatively, there exists non-oscillating solutions that appear as an ever-expanding universe from a single bounce  preceded by a collapse from the infinite past. These ever-expanding solutions provide marginally better fits to the supernova redshift data, but require larger matter densities and positive dark energy along with a positive spatial curvature.     
A qualitative analysis of CMB power spectrum in the modified theory suggests a significant negative spatial curvature, which is in stark contrast to a near-zero curvature in the standard big bang theory. 
An independent constraint on the spatial curvature can further shed light on discriminating the ever expanding and oscillatory universe scenarios.

\end{abstract}

\maketitle{}

 \section{Introduction}

The standard model of cosmology has now been accepted  to be $\LCDM$, wherein the space-time evolves according to equations of General Relativity (GR) with approximately $73 \%$ dark energy (cosmological constant) and  $27\% $ cold matter ($85 \%$ of which is invisible and possibly non-baryonic dark matter) on a more or less flat spatial geometry ($\Omega_{\Lambda} \simeq 0.73, \, \Omega_M \simeq 0.27, \, \Omega_k \simeq 0$) \cite{ade2016planck}. This standard model implies that the universe has a singular point of creation, the big-bang, at about 14 billion years ago; however since classical GR breaks down at the singularity it is believed that a complete theory of quantum-gravity would resolve the singularity  issues and explain the apparent creation of the universe. In this paper, we shall take the contrasting perspective that quantum-gravity is not essential to resolve the issues of singularity, rather a modified theory of gravity that is singularity-free at the classical level can describe the dynamics underlying the apparent `creation' process. In this paper we analyze a modified theory of gravity that adapts \emph{Torsion} to \emph{Hide} an \emph{Extra-Dimension} (THED)\cite{shankar2012metric}; it exactly matches GR  at the kinematical level, while the modifications in its dynamical equations leads to the possibility of averting classical singularities and yields eternally oscillating solutions for the universe.

\subsection{Various Approaches to Eternal Universe}

A beginning-less eternal universe has its charm because it obviates the question of creation. Soon after proposing GR, Einstein himself in 1917 contrived an unstable static universe model with a positive cosmological constant to balance the gravitational attraction of matter. With Hubble's discovery of galactic redshifts in 1929, which made it clear that the universe is currently expanding, Fred Hoyle and colleagues envisaged an expanding steady state model with a magic-mechanism of continual matter creation \cite{hoyle1948new}. After the discovery of cosmic microwave background radiation in 1965, the evolution of the universe from a hot dense state became indisputable and the big-bang model gained acceptance. The singularity issue (the creation-question) in this model has been regarded as a consequence of a lack of  quantum-gravity theory by some researchers, while some others have explored various elegant approaches to resolve the singularity issue within classical-gravity. Notable approaches include models of \emph{eternal inflation} \cite{linde1982new,albrecht1982cosmology}, \emph{emergent universe from a cosmic seed} \cite{ellis2003emergent}, \emph{bounce cosmology} \cite{creminelli2007smooth}, \emph{oscillatory universe} \cite{steinhardt2002cosmic, graham2014exploring, frampton2015cyclic},  \emph{quasi-steady-state cosmology} \cite{hoyle1993quasi}, \emph{bubble universes within blackholes} \cite{smolin1999life,poplawski2014universe}. 

In the  \emph{eternal inflation} scenario \cite{linde1982new,albrecht1982cosmology},  the universe as a whole is always inflating  while small patches of it stop inflating and reheats to fill the observable universe with matter and radiation.The singularity theorems \cite{hawking1973large} imply the existence of a singularity in the past as long as certain energy conditions \cite{visser1999energy} (NEC-Null Energy condition for spatially open or flat universes, and Strong energy condition for spatially closed universe) are satisfied. More generally, it has been shown that if the average expansion rate of the universe is positive, then irrespective of any energy condition or even the underlying theory of gravity, there must exist a singularity in the past (BGV theorem)\cite{borde2003inflationary}. This has essentially shut the door for the possibility of inflationary models to resolve the big bang singularity issue at the classical level. 

In the \emph{bubble universe inside black holes} scenario \cite{smolin1999life}, every black hole formed out of a  gravitational collapse gives birth to a baby-universe (possibly with different fundamental physical constants). For example, as matter collapses to extremely high densities inside a black hole, the torsional field coupled to the fermions  \cite{hehl1976general} is expected to generate a repulsive force and a bounce to form a baby universe \cite{poplawski2014universe}. 
However, the BGV theorem \cite{borde2003inflationary} appears to be applicable in this scenario because the expansion rate of any matter in the universe (before it starts collapsing to form the black hole) would vastly override its contraction rate during the collapse, leading us to expect a classical singularity in the past. This scenario can at best be considered an intriguing conjecture that needs quantum-gravity to resolve the details. 

In the \emph{emergent universe} scenario, the universe is a fine-tuned cosmic seed that is static in the asymptotic past, and some \emph{ad hoc} mechanism triggers its inflation and expansion.  In general, exotic-matter that violate NEC is required to contrive such a scenario \cite{mukherjee2006emergent}. However in a spatially closed universe, the cosmic seed can be tailored to be an Einstein static universe without requiring exotic-matter, but extreme fine tuning of a scalar field potential and its kinetic energy is required \cite{ellis2003emergent}. Such a configuration would be susceptible to quantum fluctuations (particularly the homogenous mode) in the field and it is highly questionable if the universe can remain asymptotically static for infinite time in the past.    

In \emph{bounce cosmology} scenario \cite{creminelli2007smooth}, the present era of expansion of the universe is preceded by a contracting phase all the way from the infinite past. Although no fine-tuning is required, an \emph{ad hoc} bounce mechanism with a ghost condensate field \cite{arkani2004ghost} that violates NEC  has to be introduced to smoothly transition from the contraction to expansion phase. At a philosophical level, it is hard to motivate the conditions wherefrom an infinitely dilute universe would homogeneously collapse.
\footnote{In both \emph{emergent} and \emph{bounce} scenarios, the universe would be in a life-friendly state (ability to energetically host life) very briefly when compared to its eternal lifetime, while in the \emph{oscillatory} scenario the universe would repeatedly spend significant fraction of time in life-friendly state. }

In the \emph{oscillatory universe} scenarios, the universe repeatedly goes through cycles of expansion and  contracting phases. Tolman \cite{tolman1931theoretical} applied the second law of thermodynamics to the universe as a whole and concluded (irrespective of the underlying theory of gravity) that the entropy of the entire universe would have to increase in every successive cycle leading to a thermal-death of the universe in a finite time, else the volume of the universe should grow in every successive cycle. 
 The Quasi-steady state model \cite{hoyle1993quasi} enforces this argument by explicitly invoking a traceless scalar field to  continuously create new matter in the universe, thereby increasing the volume and entropy of the universe in each cycle.
Some oscillatory models deal with the issue of entropy-buildup by requiring the expansion phase to be so large that all the matter debris are diluted to the extent that the universe comes back to a pristine vacuum state when it enters the contracting phase. 
For example the Baum-Frampton model \cite{baum2007turnaround, frampton2015cyclic} proposes that only a small patch of the universe comes back to the contracting phase after having jettisoned all the entropy it holds, without prescribing the actual mechanism underlying the bounce and turnaround of universe. In Steinhard-Turok model \cite{steinhardt2002cosmic}, a scalar field couples to the matter fields in a unique fashion and rolls over a tailored potential well (that acts as dark energy) to generate the expansion, turnaround and contraction of the universe. To induce a bounce from contraction back to expansion, the scalar field gets magically reflected reversing its momentum to produce some matter and radiation as byproducts for the next cycle (which is interpreted as inelastic collision between branes in higher dimensions).
All these models involve some kind of unknown mechanism that is not obviously justifiable. Furthermore, since these models are in accord with the Tolman argument, the universe cannot oscillate within fixed bounds--the minimum size of the universe increases every cycle leading to a net positive expansion rate, and application of the BGV theorem \cite{borde2003inflationary} implies an unavoidable singularity in the past. 

\subsection{ Bounded Oscillating universe}

For  models wherein oscillations happen within fixed bounds,  eternal oscillations are admissible because the BGV theorem \cite{borde2003inflationary} does not apply; however the Tolman argument \cite{tolman1931theoretical} has to be discarded. From a fundamental statistical-mechanics perspective, there does not exist an unambiguous concept of  \emph{equilibrium} or \emph{entropy} for an unconstrained isolated  system. The universe is an isolated system, its volume is dynamically unconstrained, and it is in no sense equilibrating with an external thermal bath. It may be acceptable to consider a small patch of the universe as being immersed in a thermal bath  constituted by the rest of the universe (and thereby allude to thermodynamic concepts of temperature and entropy), but there is no justification to impose the second law of thermodynamics to the universe as a whole when gravitational dynamics is involved. 
The Tolman argument has been considered a crucial theoretical (hypothetical) hindrance against the possibility of bounded oscillatory models, but in this paper we shall not adhere to that conventional viewpoint. 

All known matter fields in universe seem to satisfy the averaged NEC \cite{curiel2014primer}, so it's violation is generally considered unphysical (see \cite{rubakov2014null} for a review on NEC violations). In General Relativity  the singularity theorems prohibit a spatially flat or open universe from smoothly oscillating between fixed bounds when the NEC is satisfied.  While a closed universe can exhibit  smooth stable oscillations within fixed bounds without violating NEC \cite{graham2014exploring}, it requires abnormal-matter with negative pressure. This implies normal-matter observed in present state of universe cannot be accounted for in an oscillating universe scenario within General Relativity.  Contrastingly, in modified theories of gravity like THED \cite{shankar2012metric}, bounded oscillating universe with normal-matter (like dust)  is permissible with any spatial topology.

\subsection{Organization of the paper}
 
In section 2, we start with a review of THED and then present the cosmological equations guiding the evolution of the universe. In section 3, the oscillatory universe solutions are analyzed  and the parameter space demarcating the ever-expanding and oscillatory solutions is discussed. Importantly, it is shown that the overall energy of the universe within each oscillation is precisely zero. 
In section 4, we use the Type Ia  supernovae redshift data from an open source catalog \cite{SupernovaDataset} (https://sne.space/) and fit it to evaluate the cosmological solutions and extract the best fit parameter values.    
In section 5, we analyze the implication of the cosmic microwave background (CMB) power spectrum \cite{spergel2003first} and infer that it suggests a negative spatial curvature for the universe, which is in stark contrast to the requirement of a near-zero curvature in $\LCDM$ model.
 In section 6, we discuss the implications of the results  and the amicability of  THED as a theory of modified gravity. 

\section{Reviewing $THED$ gravity}

\emph{Torsion} is the natural mathematical ingredient of differential geometry that can be incorporated into the spacetime manifold (in addition to its \emph{metric} structure) in order to include the gravitational effects of matter with spin \cite{hehl1976general}. However, torsion plays a completely different role in THED gravity \cite{ shankar2012metric}-- here torsion is not coupled to  the spin of the matter fields, instead it serves to hide an extra-dimension in space.  A brief technical review of this theory follows. The reader can skip this review and move to the cosmological equations without loosing coherence. 

We start by visualizing the 5D space-time as  foliated  4D hypersurfaces (with coordinates $x^{\mu}$) along the fifth dimension $x^{5}$.
\begin{eqnarray}
ds_{5}^2 &=& \g_{\mu \nu}  dx^{\mu} dx^{\nu} + \g_{\mu 5} dx^{\mu} dx^{5} + \Phi^{2} [dx^5 ]^2  \nonumber \\
&+& \g_{\mu 5} \g_{\nu 5} \Phi^{-2} dx^{\mu} dx^{\nu} . 
\label{5Dmetric}
\end{eqnarray}
The line element is expressed in the above form for convenience because once we impose the physical constraint that the extra-dimension remains hidden, it turns out that the 4D components $\g_{\mu \nu}$ can be exactly identified as the metric tensor in torsion-free GR at a kinematical level.

The physical constraint is laid on the 5D geodesic equation such that any motion along the 5th dimension should have no effect on the observable 4D motion. That is, irrespective of the velocity along the 5th coordinate, the 4D equations describing the motion is required to be invariant--thus effectively hiding the existence of the 5th dimension. Mathematically, this leads to constraining the 5D connection $\Christilde{}{}{}$, namely  $\Christilde{\mu}{\nu}{5} = \Christilde{\mu}{5}{5}=0$, and the requirement of torsionless 4D geometry leads to $\Christilde{\mu}{[ \alpha}{\beta ]}=0 $.
A vierbein formulation of these constraints is elaborated in  \cite{shankar2010kaluza}.  These constraints along with the metric-compatibility condition completely determine all the components of torsion in terms of the metric. 

Remarkably, this theory is indistinguishable from GR at a kinematic level. First, it turns out that the 4D metric on the hypersurfaces, $\g_{\mu \nu}$, is functionally independent of $x^{5}$--making all the hypersurfaces identical; this can be viewed as a strong validation to our procedure of hiding the extra-dimension, and to set the cylindrical condition (meaning, the entire 5D metric is independent of $x^{5}$). Second, the components of the connection and the Ricci tensor that are tangential to any hypersurface are identical to those in  GR with metric $\g_{\mu \nu}$. Mathematically, this implies that the terms contributed by the torsion exactly cancel-off the terms contributed by the effect of the extra-dimension in the computation of the 4D components of connection and Ricci tensor.  Hence, hiding the extra-dimension by requiring that the physically observable 4D motion is oblivious to any extra-dimensional motion, leads to a theory where the kinematic equations are identical to those in GR.  Moreover, this formalism does not require us to explicitly choose a topology for the extra dimension--it could be either a compact dimension or a large unbounded dimension. 
 
Since torsion is not dynamically independent and is completely metric dependent, the field equations are derived from the action principle by varying just the metric, and they take the form,\footnote{ In the original derivation of eq.~\ref{FE} presented in \cite{shankar2012metric}, the sign convention adopted to define the curvature and the Einstein tensor is different.}
\begin{eqnarray}
G^{\mu}_{\nu} -  \h^{\mu}_{\nu}=   \Sigma^{\mu}_{\nu} ,
\label{FE}
\\
\g^{\mu 5} [ \R_{\mu \nu} - \h_{\mu \nu}] = \Sigma^{5}_{\nu}  , \,\,  \frac{1}{2} \R=\Sigma^{5}_{5}. 
\end{eqnarray} 
Here $\Sigma$ is the 5D stress tensor with  $\Sigma^{\mu}_{\nu}$ being the physically observable 4D components.
$G^{\mu}_{\nu}$ is the standard Einstein tensor constructed from the 4D metric $\g_{\mu \nu}$ as in  GR, and $\R_{\mu \nu}$ and $\R$ are respectively the  4D Ricci tensor and Ricci scalar  constructed from the 4D metric $\g_{\mu \nu}$ as in GR. The additional term $\h^{\mu}_{\nu}$ (under cylindrical condition) takes the form
\begin{equation}
\h^{\mu}_{\nu} =  \nabla_{\nu} \J^{\mu} - (\nabla . \J) \delta^{\mu}_{\nu} 
  + \J_{\nu} \J^{\mu}- (\J.\J)\delta^{\mu}_{\nu},
  \label{defH}
\end{equation}
where $\J_{\mu} \equiv  \Phi^{-1} \partial_{\mu} \Phi $, is  a 4 vector whose indices are raised and lowered with the 4D metric $\g_{\mu \nu}$ and its inverse. Similarly, the covariant derivative,  $\nabla_{\nu}$, is defined as in GR with Christoffel connection expressed in terms of the 4D metric $\g_{\mu \nu}$.
Note that the 4D components of the field equations only involve the 4D metric components $\g_{\mu \nu}$ and the scalar function $\Phi$ (which is simply $\g_{55}$), while the metric components $\g_{\mu 5}$ have completely decoupled out. 
Particularly, $\g_{\mu 5} =0$ and $\Sigma^{5}_{\nu}=0$ are always consistent solutions to the field equations, but there is absolutely no necessity to impose them.  Since only the 4D metric components $\g_{\mu \nu}$ and the 4D components of stress tensor $\Sigma^{\mu}_{\nu}$ are relevant for physical observations, eq.~\ref{FE} is sufficient to solve for all the physical degrees of freedom including $\Phi$. The unobservable extra dimensional stress tensor component $\Sigma^{5}_{5}$ can be defined to be $\mathcal{R}/2$ consistent with the field equations.

Clearly, this theory reduces to GR when $\J_{\mu}$ vanishes. In GR, the Bianchi identity automatically implies the conservation of stress tensor, which is not true here because of the presence of the term $\h^{\mu}_{\nu}$ in eq.~\ref{FE}. Nevertheless, if we impose 4D matter conservation, $\nabla_{\mu}  \Sigma^{\mu}_{\nu} =0$, it would imply $\nabla_{\mu}  \h^{\mu}_{\nu} =0$. Moreover, since this condition must hold in vacuum regions of spacetime, and since there is no reason for a purely geometry-dependent tensor  $\h^{\mu}_{\nu}$ to behave differently in vacuum \emph{vs} matter-filled regions, we would expect any smooth solution of field equations (eq.~\ref{FE}) to automatically satisfy $\nabla_{\mu}  \h^{\mu}_{\nu} =0$. 

 Finally, notice that the field equations (eq.~\ref{FE}) are second order differential equations when they are expressed in terms of the dynamical variables (metric components), and hence we do not have to worry about Ostrogradsky instabilities \cite{woodard2015theorem} in the solutions.

\subsection{Interpretation of the extra-dimension}
 
A natural question that emerges is how particles at different $x^{5}$ positions (but at same physical 4D position) would interact with each other. Would they bypass each other or  would their interaction be guided by the presence of the hidden stress tensor component $\Sigma^{5}_{5}$? These questions can be avoided if all physical particles are constrained to a single hypersurface with a fixed $x^{5}$ coordinate, but that would defeat the primary purpose of hiding the 5th dimension. 

Moreover, since the metric is independent of $x^{5}$, the stress tensor should also be independent of $x^{5}$. Hence the matter distribution must be identical on all the 4D hypersurfaces. That is, if a particle exists at a particular 4D coordinate on a hypersurface, then such a particle should exist on every hypersurface at that 4D coordinate. In other words, every particle in 4D is essentially a \emph{string} in 5D.  The physical constraints imposed on the geodesics would then imply that any motion of the string (extended along $x^{5}$) along itself would have no effect on the equations of motion of the 4D coordinates. As these strings move around in the 4D coordinates, their interaction would be identical to the interaction of particles in torsion-free 4D spacetime of GR
\footnote{We could potentially speculate subtle new-physics in the interactions that depend on the unobservable internal motion of the strings along $x^{5}$ }.
Thus the interpretation that the particles in the observable 4D spacetime are extended strings along the hidden fifth dimension is a natural outcome of the hiding-constraints imposed on the geodesics. However, this interpretation has no bearing on the rest of this paper  because we only focus on the observable 4D field equations (eq.~\ref{FE}) and analyze them in the light of cosmological data.

\subsection{Cosmological equations}

For homogeneous and isotropic universe, the 4D line element in spherical-polar coordinates is given by 
\begin{equation}
ds_{4}^{2} = - dt^{2} +a^{2}(t) \left( \frac{dr^2}{1-kr^2} +r^2 d\Omega^{2}  \right), 
\label{4Dmetric}
\end{equation}
where $a(t)$ is the scale factor, and  $k=0$  corresponds to spatially flat universe, $k>0$ corresponds to a spatially closed universe, and $k<0$ corresponds to spatially open universe. The nonvanishing components of the Einstein tensor for the above metric is given by 
\begin{eqnarray}
-\mathrm{G}^{t}_{t} &=& 3 (\dot{a}/a)^{2}  + 3 k/a^2 , \nonumber \\
-\mathrm{G}^{r}_{r} &=&  2 (\overset{..}{a}/a) + (\dot{a}/a)^2 + k/a^2 , \nonumber \\
 \mathrm{G}^{\theta}_{\theta} &=& \mathrm{G}^{\phi}_{\phi} = \mathrm{G}^{r}_{r} ,
 \label{CosmG}
\end{eqnarray}
where  over-dot denotes a derivative with respect to time. With the 4D metric (eq.~\ref{4Dmetric}) embedded in the 5D geometry (eq.~\ref{5Dmetric}), and acknowledging that the extra-dimensional metric field $\Phi$ can functionally depend only  on time $t$, the only nonvanishing component of $\J_{\mu}$ is  $\J_{t}$ and the nonvanishing components of $\h^{\mu}_{\nu}$  in  eq.~\ref{defH} are
\begin{eqnarray}
\h^{ t}_{t}  &=& 3\J_{t} (\dot{a}/a) ,  \nonumber \\
 \h^{ r}_{r} &=& 2 \J_{t} (\dot{a}/a) +  \dot{\J}_{t} +  \J^{2}_{t} , \nonumber \\
 \h^{ \theta}_{\theta}  &=& \h^{ \phi}_{\phi} = \h^{ r}_{r} .
\end{eqnarray}
The conservation equation $\nabla_{\mu}  \h^{\mu}_{\nu} =0$ implies that either $\J_{t}=0$ or $\J_{t}=  \overset{..}{a}/\overset{.}{a}$. That is, either $\Phi(t)$ is a constant, which would give rise to the usual Friedman-Robertson-Walker (FRW) cosmology, or $\Phi(t)= \dot{a}(t)$. Focusing on the latter case, $\h^{\mu}_{\nu}$ simplifies to 
\begin{equation}
\h^{ t}_{t} = 3 \overset{..}{a}/a, \qquad  
\h^{ r}_{r} =  2 (\overset{..}{a}/a)  +  (\overset{...}{a}/\dot{a}).
\label{CosmH}
\end{equation}
and the field equations (eq.~\ref{FE}) reduce to
\begin{eqnarray}
 3 (\dot{a}/a)^{2}  + 3 k/a^2 +  3 \overset{..}{a}/a  &=&  8 \pi G \rho. 
  \label{eqmRho}  \\
 4 (\overset{..}{a}/a)   + (\dot{a}/a)^2 + ( \overset{...}{a}/\dot{a})  + k/a^2  &=& -  8 \pi G P . 
 \label{eqmP}
  \end{eqnarray}
Here $\rho$ and $P$ are the density and pressure of the 4D matter, and $G$ is the Newton's gravitational constant. 

The existence of the third order derivative $\overset{...}{a}$ in eq.~\ref{eqmP} might at first glance provoke a suspicion indicating Ostrogradsky instability \cite{woodard2015theorem}.  But this is because the conservation equation has only been partially utilized to rewrite a dynamical variable ($\Phi$) in terms of the derivative of another dynamical variable ($a$). However, when the conservation equation is fully utilized, the third derivative will no longer exist in the field equations and hence there are no threats of instabilities in the field equations. To complete the employment of conservation equation, we look at the matter components and impose the matter conservation equation $\nabla_{\mu}  \Sigma^{\mu}_{\nu} =0$ to obtain
\begin{equation}
\dot{\rho} + 3(\rho + P) \dot{a}/a =0 ,
\label{cons}
\end{equation}  
which exactly resembles the conservation equation in  FRW cosmology. Now eq.~\ref{eqmRho} and eq.~\ref{cons} together represent the complete set of field equations making eq.~\ref{eqmP} and it's third derivative term redundant.   

For the standard  matter equation of state, $P= w \rho$, it is well known from solving eq.~\ref{cons}  that the density evolves with time as $\rho(t) \propto a(t)^{-3(1+w)}$. The Null Energy condition (NEC) is satisfied if $\rho +P \ge 0$, which is the kind of matter we observe in the universe.  

To keep the analysis simple, we shall just consider normal pressure-less matter $(w=0)$ with density $\rho_m$ and a cosmological constant $(w=-1)$ of density $\Lambda$. The NEC is satisfied if $\rho_m \ge 0$  irrespective of the sign of $\Lambda$. To solve the cosmological equations (eq.~\ref{eqmRho}, \, \ref{cons}), we shall set the initial conditions at the present moment ($t=0$) to be such that $a(t=0)=1$ and $\dot{a}(t=0)=1$, assuming that the universe is currently expanding. This sets the units of length and time scales in our numerical computations to be such that the Hubble constant $H_o \equiv [\dot{a}/a]_{t=0}$ and the velocity of light $c$ are taken to be 1. This implies the physical time and physical length should be numerically interpreted in the units of $H_o^{-1}$ and  $cH_o^{-1}$ respectively. 

The equation to solve is then
\begin{eqnarray}
\mathrm{THED}:    \left[ \overset{..}{a}/a \right] +(\dot{a}/a)^{2}  - \Omega_{k}/a^2 &=&   \Omega_{M}/a^3 + \Omega_{\Lambda} 
\label{cosm} \\
\mathrm{GR }:  \qquad \,\,\,\,\,\,\,\,\,\,(\dot{a}/a)^{2}  - \Omega_{k}/a^2  &=&   \Omega_{M}/a^3 + \Omega_{\Lambda}
\label{cosmFRW}
\end{eqnarray}
In effect, the  GR field equations are modified in THED gravity by the inclusion of the first term in l.h.s of eq.~\ref{cosm}, namely $\left[ \overset{..}{a}/a \right] $.  
$\{ \Omega_k,  \Omega_{M},\Omega_{\Lambda} \} $ are density parameters (energy density divided by critical density) at the present moment $t=0$. 
In GR $\Omega_M + \Omega_{\Lambda} + \Omega_k =1$, while in THED $\Omega_M + \Omega_{\Lambda} + \Omega_k =1 -q$, where deceleration parameter  $q \equiv [- \overset{..}{a} a/\dot{a}^2]_{t=0}$. 

\textbf{Physical values of the energy densities :}
The physical value of $\Omega_k$ is given by $\Omega_k=-kc^2/H_o^2$. Notice that $k$ and $\Omega_k$ have opposite signs, so a closed universe with a positive curvature will have a negative $\Omega_k$ and an open universe with a negative curvature will have a positive $\Omega_k$.
The physical density of matter  is $(3H_o^2/8 \pi G) \Omega_M$ and the  dark energy density  is $(3H_o^2/8 \pi G) \Omega_{\Lambda}$.  For example, if $H_o=70$ km/s/Mpc, the physical matter density turns out to be $ \Omega_M *0.9 * 10^{-26} kg/m^3 $.  The observed matter density of luminous gasses in the universe only amounts to a contribution of $\Omega_M \simeq 0.005$; however with theoretical extrapolations from bigbang neucleosynthesis the total baryonic matter can amount to $\Omega_M \simeq 0.05$, while the standard model $\LCDM$ requires $\Omega_M \simeq 0.27$ including dark matter contributions. For completion, we should ideally include a radiation density term of $\Omega_R /a^4$ in the r.h.s of eq.~\ref{cosm}. But since observations reveal that CMB radiation $\Omega_R \simeq 5 * 10^{-5}$ is extremely small in comparison to matter density, its effect will be significant only when  $a < 10^{-4}$ and so it can be ignored while studying the evolution history of the universe at larger scales. 

\section{Oscillatory Universe solutions}
\label{profile}

\begin{figure}
\begin{center}
\includegraphics[width=0.4\textwidth]{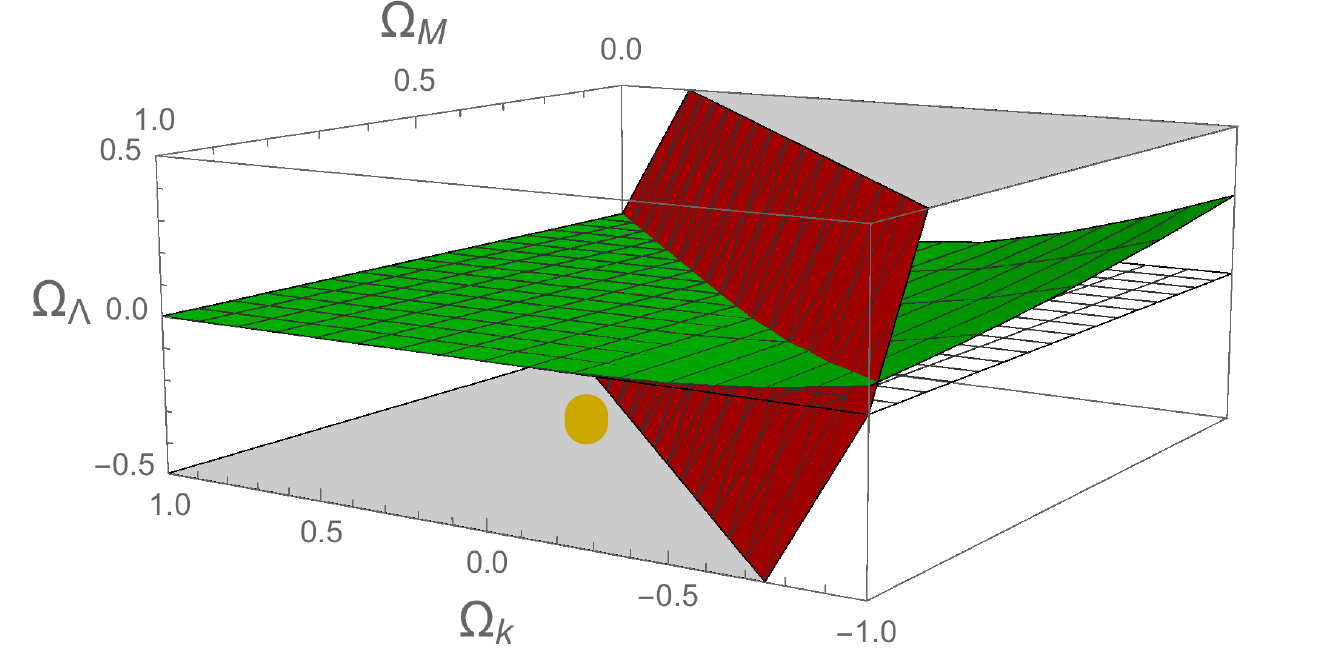} 
\end{center}
\caption{The parameter space  $\{ \Omega_k,  \Omega_{M} ,\Omega_{\Lambda} \} $  is classified into regions that accommodate solutions of universe that (i) bounces from a minimum nonzero size and (ii) turns around from an expansion phase to a contraction phase at some maximum size.  The flat transparent horizontal surface marks the $\Omega_{\Lambda}=0$. The  green-surface (nearly horizontal) marks the boundary below which an expanding universe turns around at a maximum and starts contracting, while above the surface an expanding universe will continue to expand forever.  The red-surface  (nearly vertical) marks the boundary to the left of which a contracting universe bounces back to expand from a minimum size. The bottom-left quadrant  of the parameter space (marked by the brown dot) supports bounded oscillating universe. For visual clarity, the values of $\Omega_{\Lambda}$ is restricted between $\pm 0.5$. }
\label{paramfigure}
\end{figure}

Oscillatory solutions of eq.~\ref{cosm} exist when the parameters   $\{ \Omega_k,  \Omega_{M},\Omega_{\Lambda} \} $ lie in a particular region of the parameter space, as depicted in figure~\ref{paramfigure}. For the expanding universe to reach a maximum size $a_{max}$ and turn around to contract, the acceleration at the maximum should be negative. 
\[ \overset{..}{a}_{max} = \Omega_k /a_{max} + \Omega_{\Lambda} a_{max} +\Omega_M / a_{max}^2 <0 \]
Similarly, for the universe to bounce back from a minimum size $a_{min}$, the acceleration at the minima should be positive.
 \[ \overset{..}{a}_{min} = \Omega_k / a_{min} + \Omega_{\Lambda} a_{min} +\Omega_M /a_{min}^2  >0 \]

However, $a_{max}$ and $a_{min}$ are themselves implicitly determined by the parameters through eq.~\ref{cosm}. For the universe to have a nonsingular bounce, the parameters have to lie to the left of the red-surface in figure \ref{paramfigure}. For the universe to turn around from a maximum, the parameters should lie below the green-surface in figure \ref{paramfigure}. Effectively, the parameters corresponding to the oscillatory solutions lie in the bottom-left quadrant of figure \ref{paramfigure} as indicated by a brown dot. 

 \begin{figure*}
\includegraphics[width=0.32\textwidth]{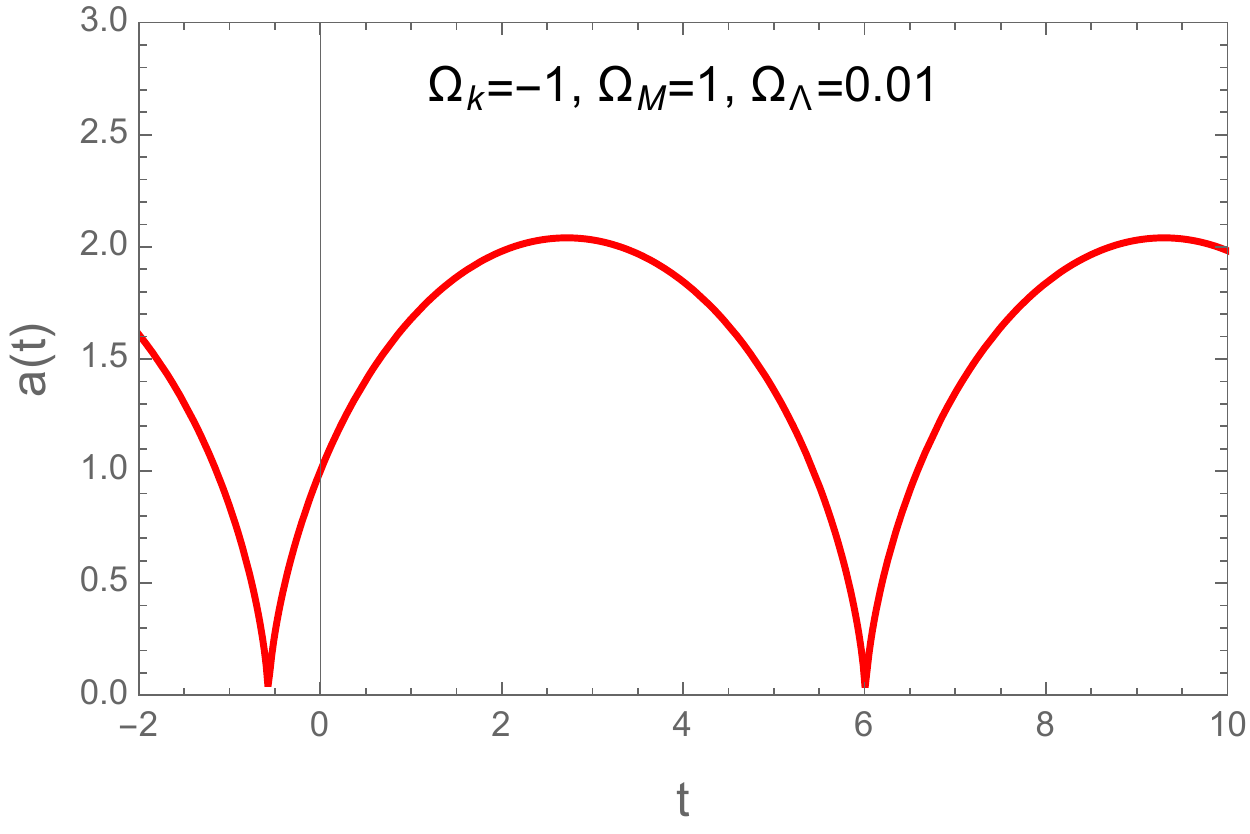} 
\includegraphics[width=0.32\textwidth]{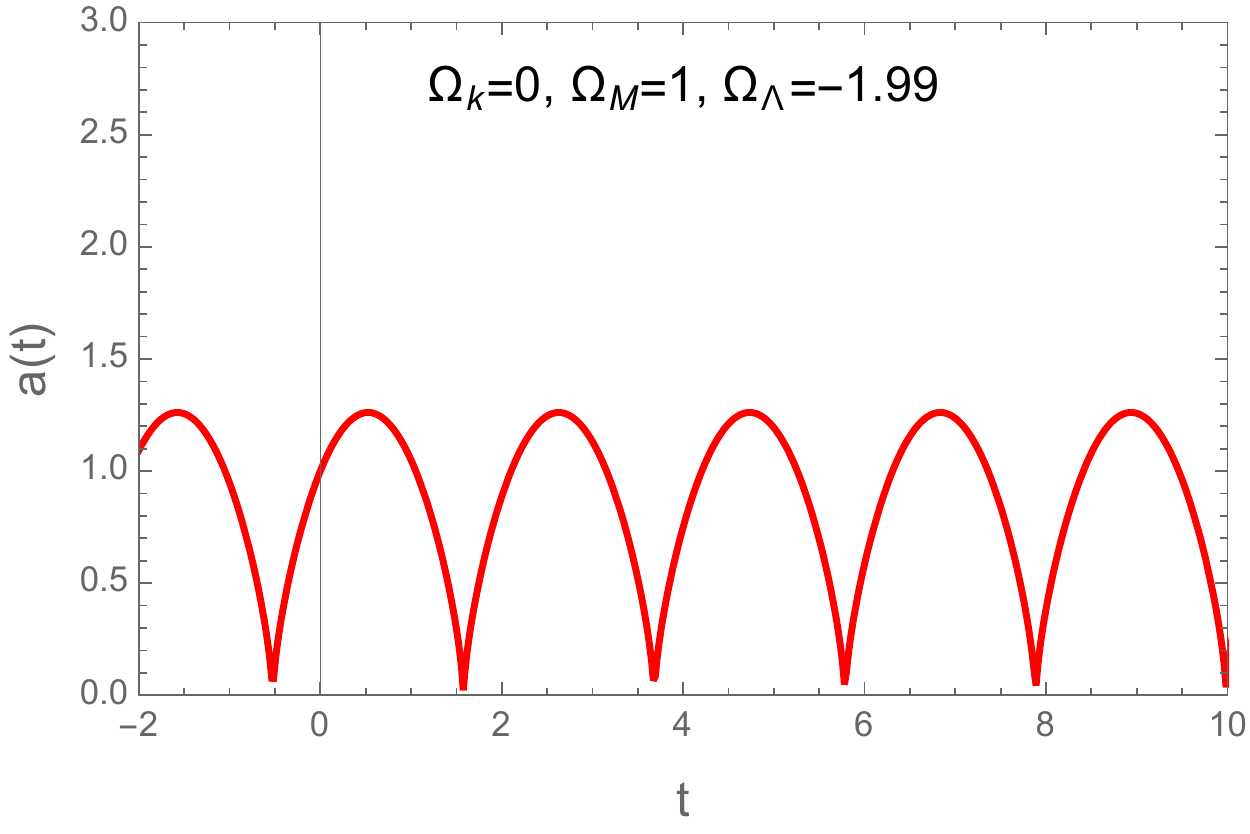}
\includegraphics[width=0.32\textwidth]{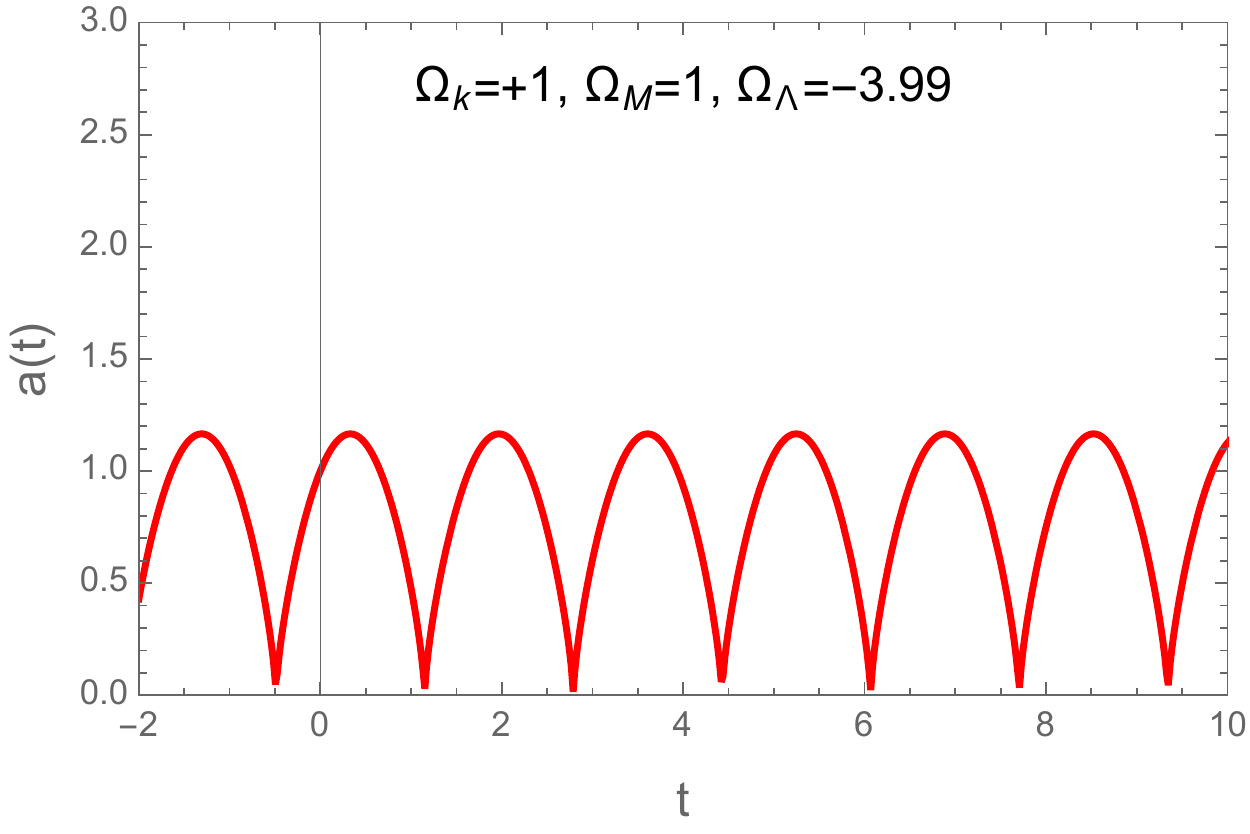}
\\  \qquad \\
\includegraphics[width=0.32\textwidth]{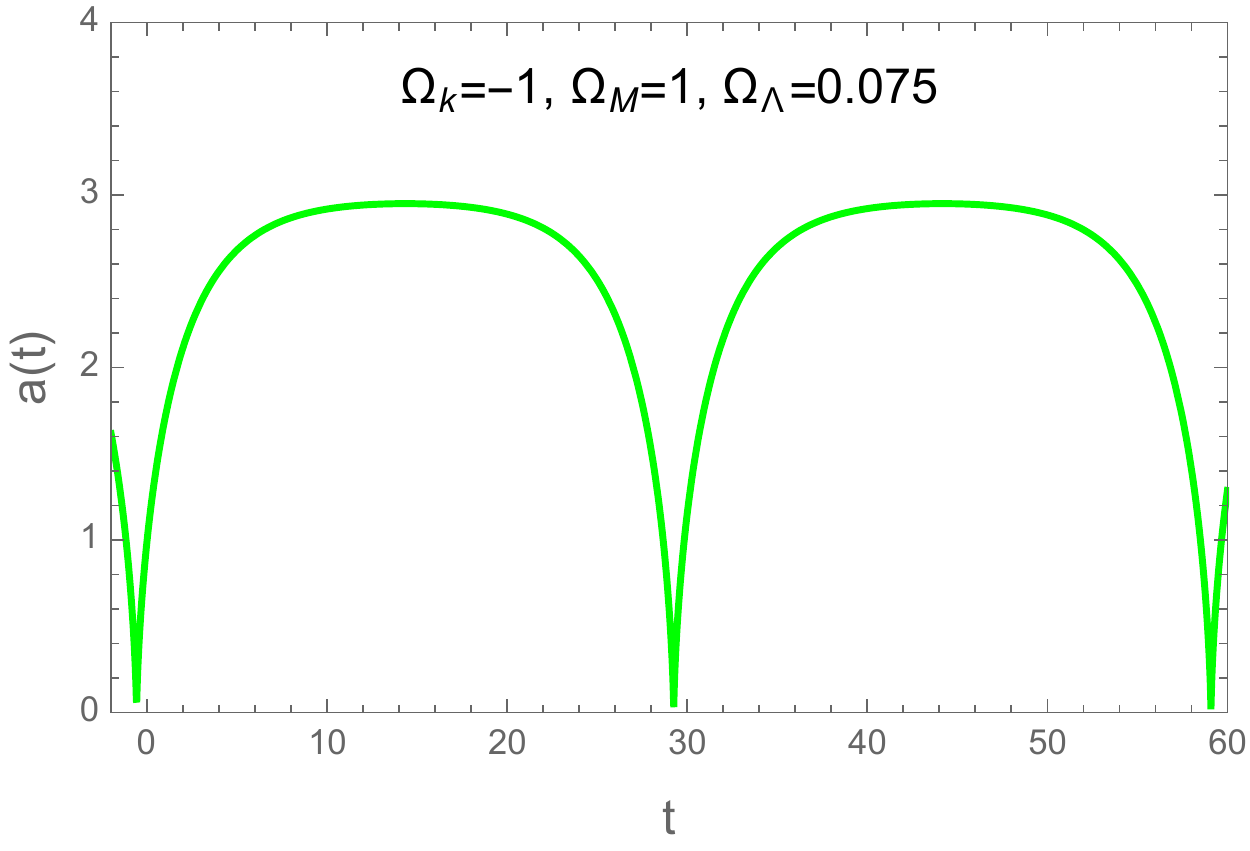} 
\includegraphics[width=0.32\textwidth]{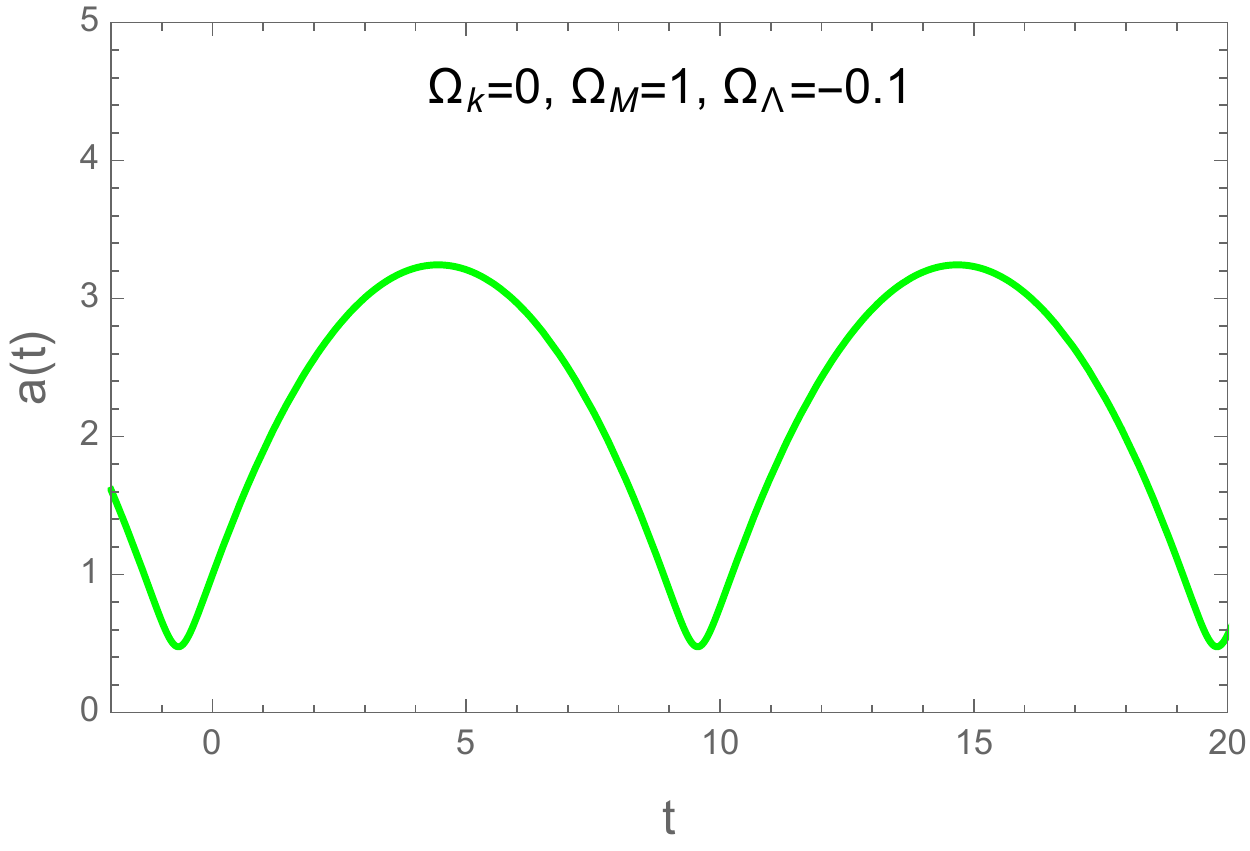}
\includegraphics[width=0.32\textwidth]{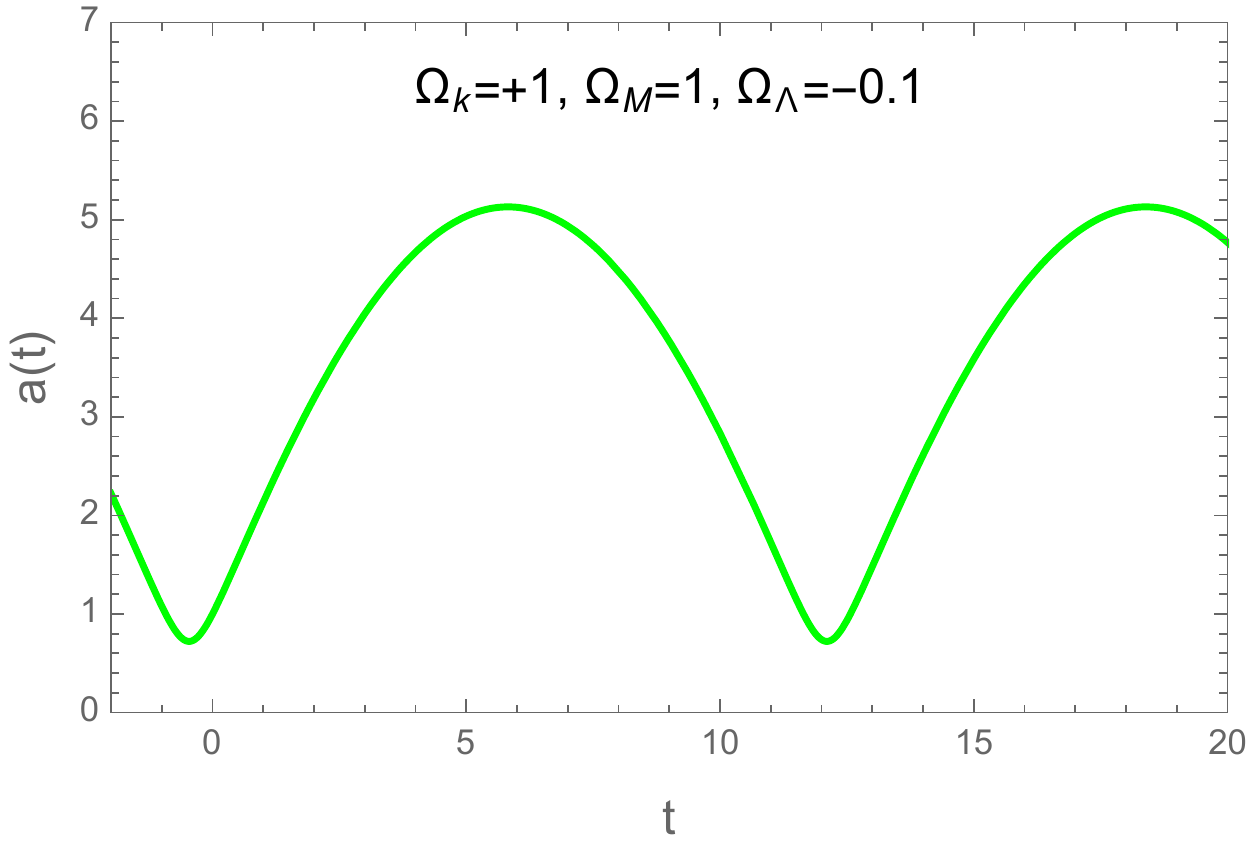}
\\ \quad \\
\caption{Illustration of oscillatory solutions of $a(t)$ for three pairs of $ \{ \Omega_k, \Omega_M  \} $. The top row corresponds to parameters lying close to the red-surface in fig.~\ref{paramfigure}, and the bottom row corresponds to the parameters lying close to the green-surface in fig.~\ref{paramfigure}.} 
\label{Osfig}
\end{figure*}

The minimum size of the universe $a_{min}$ depends on how close the parameters lie from the red-surface in figure~\ref{paramfigure}. The value of $a_{min}$ can be pushed arbitrarily close to zero by moving the parameters very close to the red-surface, which empirically  evaluates to be
\begin{equation}
\Omega_{\Lambda} \ge  \,\,\, \Omega_{\Lambda}^{r} \equiv 2 -2 \Omega_k  - 4 \Omega_{M}.
\label{redsurface}
\end{equation}
$\Omega_{\Lambda}^{r}$ is the value of $\Omega_{\Lambda}$ at the the red-surface. The exact position of the red-surface will be minutely affected by the presence of radiation density (additional $\Omega_R /a^4$  term in the r.h.s of eq.~\ref{cosm}). For example,  consider a point on the red-surface with $ \{\Omega_k = 0.9$, $\Omega_{\Lambda} =0$, $\Omega_M =0.05 \}$ evaluated with the assumption that $\Omega_R =0$.   By including the currently observed $\Omega_R =5 * 10^{-5}$, that point shifts to $\{ \Omega_k = 0.9$, $\Omega_{\Lambda}=0$, $\Omega_M=0.0494 \}$. In other words, including the radiation density has the effect of minutely decreasing the value of $\Omega_M$ on the red-surface given by eq.~\ref{redsurface}, but this does not meaningfully alter the characterization depicted by figure \ref{paramfigure}. 

Although we can ensure  $a_{min}$ to be extremely small, its exact value depends on the fine-balance between the parameters in eq.~\ref{redsurface}  (cannot be constrained by any observation). Hence for all practical purposes, we can treat $a_{min}$ to be independent after constraining the parameters to the red-surface of figure~\ref{paramfigure}.

It can be straightforwardly deduced that an oscillatory universe with $a_{min} \rightarrow 0$ should not be accelerating at the current epoch. Note that with $\Omega_k>0$, we need $\Omega_{\Lambda}$ to be negative and also satisfy eq.~\ref{redsurface}. The deceleration parameter given by $q=1-[\Omega_{\Lambda}+\Omega_{k}+\Omega_{M}]$ then simplifies to $q \simeq \Omega_M - \Omega_{\Lambda}/2$ which has to be positive. The value of $q$ can be negative only for parameters that lie far away from the red-surface or lie above the green-surface in figure~\ref{paramfigure}.

The periodicity of oscillations  $\tau$ is determined by how far away the parameters are from the green-surface. The closer the parameters get to the green-surface, $\tau$ increases without bounds.  Thus the parameter values close to the intersection of the red and green surfaces will correspond to  solutions with  $a_{min} \rightarrow 0$ and $\tau \rightarrow \infty$. 
For $\Omega_k \ge 0$ (spatially flat/open universe), the green-surface is given by $\Omega_{\Lambda} =0$, while for  $\Omega_k < 0$ (spatially closed universe), the green-surface has $\Omega_{\Lambda} >0$ (see figure~\ref{paramfigure}). As the parameters get close to the green-surface the value of $a_{max}$ and $\tau$ both grow unbounded for $\Omega_k \ge 0$. But  for $\Omega_k <0$, the value of $a_{max}$  remains bounded while $\tau$  grows unbounded as we get close to the green-surface.

Figure ~\ref{Osfig} illustrates the influence of the parameters on the oscillatory profile and the values of $a_{min}$, $a_{max}$ and $\tau$. For simplicity we fix $\Omega_M=1$ for three different values of $\Omega_k = 0, +1,-1$. Consider two values of $\Omega_{\Lambda}$, one close to the red-surface and the other close to the green-surface. Imagine the brown dot in figure~\ref{paramfigure} lying close to the red-surface, and then imagine it to be vertically transported towards the green-surface without changing the values of $\Omega_k$ and $\Omega_M$. The top row in  figure~\ref{Osfig} shows the oscillations when the parameters lie close to the red-surface in figure~\ref{paramfigure}, and the bottom row shows the oscillations when the parameters lie close to the green-surface. All relevant properties can be gleaned by comparing the top and bottom rows of figure~\ref{Osfig}.

\begin{itemize} 
 \item{
 From the top row, note that $a_{min}\simeq 0$ irrespective of the position of the parameters on the red-surface, but the value of $a_{max}$ and $\tau$ depend on the parameters. 
 }
 \item{
 As we move away from the red-surface towards the green-surface (bottom row), note that $a_{min}$, $a_{max}$ and $\tau$ all increase. It is particularly crucial to note that this happens even with $\Omega_k <0$ (wherein the green-surface lies at a positive $\Omega_{\Lambda}$), as seen in the bottom-left panel (although it is not visually discernible that $a_{min}$ has moved away from 0).  
} 

\item{
The values of $\Omega_{\Lambda}$ are chosen in order to get close to the red and green surfaces (but not exactly on the surfaces), we can nevertheless get arbitrarily close to the surfaces by fine-tuning the chosen  $\Omega_{\Lambda}$. If we fine-tuned $\Omega_{\Lambda}$to get closer to the respective surfaces, the oscillations in the top row of fig.~\ref{Osfig} would not show any visually discernible change (because $a_{min}$ is already very close to zero), but the scale of oscillations in the bottom row would be drastically affected.  
}
\item{ 
For $\Omega_k=0,+1$ in the bottom row, $a_{max}$ is finite because the choice $\Omega_{\Lambda}=-0.1$ is not on the green-surface. However, we could move much closer to the green-surface by pushing $\Omega_{\Lambda} \rightarrow 0$, and that would push $a_{max} \rightarrow \infty$ along with  $\tau \rightarrow \infty$. But for $\Omega_k=-1$ (bottom-left panel), moving arbitrarily close to the green-surface will only push $\tau \rightarrow \infty$, while $a_{max}$ will saturate and remain bounded as indicated by the extremely flat plateau (rather than a peak)----a very lengthy static-phase for the universe at the crest of the oscillation.    
 }

 \end{itemize}

The parameters thus drastically affect the oscillatory profile, however there is a constant integral of motion that evaluates to zero for any parameter values in the oscillatory regime.  This constant of motion can be interpreted as  
the average energy of the universe within each cycle. 

\subsection{Zero Energy Oscillations}

\begin{figure}
\includegraphics[width=0.4\textwidth]{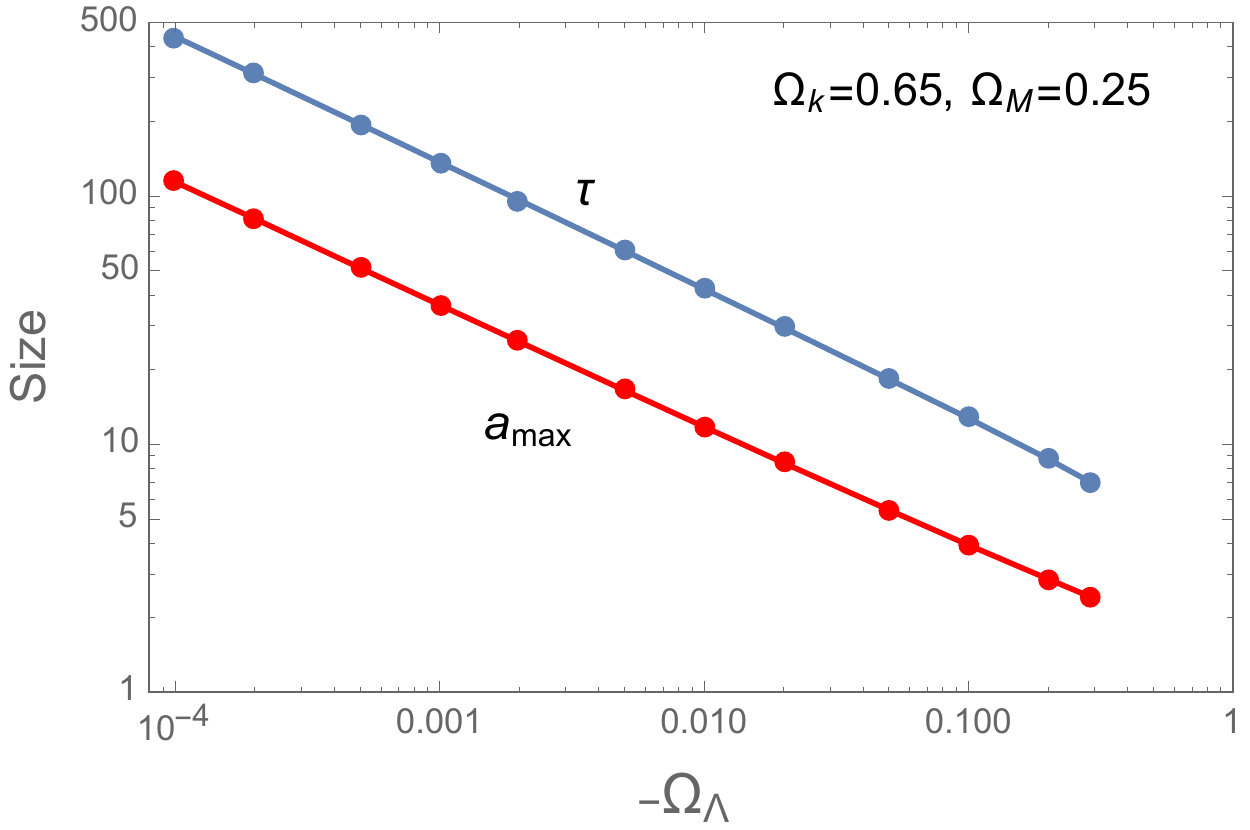} 
\caption{Periodicity $\tau$ and maximum size $a_{max}$ increase as $\Omega_{\Lambda} \rightarrow 0$ for a fixed $\Omega_M$ and $\Omega_k$. For $\Omega_k=0.65 $ and $\Omega_{M}=0.25$, the red-surface is located at the value $\Omega^R_{\Lambda} =-0.3$, hence the plot stops there. }
\label{Sizefig}
\end{figure}

The equation determining the dynamics of $a(t)$, namely eq.~\ref{cosm}, can be rewritten as
\begin{equation}
\frac{d }{dt} (a \dot{a}) = [ \Omega_M /a^3 +\Omega_k/a^2 + \Omega_{\Lambda} ] a^2 \equiv E \, a^2
\label{eq:exact}
\end{equation} 
where $E \equiv [ \Omega_M /a^3 +\Omega_k/a^2 + \Omega_{\Lambda} ] $ is the total energy density of the universe at any moment. Equivalently,
\begin{equation}
\frac{1}{2} d (a \dot{a})^2 = E \, a^3 \, \dot{a} \, dt
\end{equation} 
Since the RHS is an exact differential and since $\dot{a}$ vanishes at both the minima and maxima within an oscillation, the RHS must vanish when integrated from the minima to the maxima within a cycle.
 \begin{equation}
\int_{t_{min}}^{t_{max}} E \,a^3\, \dot{a} \, dt 
= \int_{a_{min}}^{a_{max}} E \, a^3 \,da  =0
\label{constmotion}
\end{equation} 
Thus the average energy of the universe within each half-cycle is by itself zero. This is a constant of motion, and is particularly useful to analytically evaluate $a_{max}$.  In the limit $a_{max} \gg a_{min}$, the above integral reduces to 
\begin{equation}
\Omega_M + \frac{1}{2} \Omega_k a_{max} + \frac{1}{4} \Omega_{\Lambda} a_{max}^3 = 0, 
\end{equation}  
from which $a_{max}$ can be extracted. For $\Omega_k >0$, as we approach the green surface $(-\Omega_{\Lambda} \rightarrow 0 )$, the value of $a_{max}$ and  $\tau$ are proportional to $ 1/\sqrt{-\Omega_{\Lambda}} $. This is well illustrated in fig.~\ref{Sizefig}, as the log-log plot of $a_{max}$ and $\tau$ \emph{vs} $-\Omega_{\Lambda}$ turns out to be a straight line with a slope -1/2. We can then deduce from the above equation that 
\begin{equation}
a_{max}  \simeq  \frac{\sqrt{2 \Omega_k}}{\sqrt{-\Omega_{\Lambda}} }  \qquad \textrm{as} \qquad   
\{ -\Omega_{\Lambda} \rightarrow 0 \}
\label{tiny}
\end{equation} 
It should be noted that $\Omega_M$ is implicitly fixed by eq.~\ref{redsurface} so that the parameters lie close to the red-surface of fig.~\ref{paramfigure}.

It is important to note that eq.~\ref{constmotion} does not denote the conventional sense of time-averaging over energy, rather the averaging is done over a cosmic time unit that flows in lieu with the expansion rate of the universe. This is however not the only constant of motion. We can derive other constants of motion, for example a time-averaged surface-energy in each half-cycle must also be zero. 
\begin{equation}
\int_{t_{min}}^{t_{max}}  E \, a^2 dt   \, =  \,0 
\end{equation} 
This can be interpreted as the time-averaged energy over any 2D-comoving surface in the universe to be zero.   

The existence of these constants of motion is a very unique property of the THED dynamical equation (eq.~\ref{cosm})  because  it can be written in the form of first order exact differential (eq.~\ref{eq:exact}) with a term $\dot{a}$ inside the differential. This is not possible in the usual Friedman equation (eq.~\ref{cosmFRW}) where there are no second derivative terms of $a(t)$. 

\subsection{Expansion from the Minima}

Assuming that the minimum size of the universe  $a_{min}$ is very small, the field equation (eq.~\ref{cosm}) can be solved in the neighborhood of the minima, where only the $\Omega_M$ term will dominate.  More generally,  by setting the initial conditions as $a(t=t_{min}) = a_{min}$ and $\dot{a}(t=t_{min}) =0$ at the minima, the field equations with just one dominant energy component (given by $n$) is
\begin{equation}
    \left[ \overset{..}{a}/a \right] +(\dot{a}/a)^{2}   \simeq   \Omega/a^n .
    \label{mineq}
\end{equation}    
Let $\tilde{a} = a/a_{min}$ and $\tilde{t} =(t-t_{min}) \sqrt{\Omega / a_{min}^n} $, then for $\tilde{a} \gg 1$, the equation can be asymptotically solved as
\[
\tilde{a}  \simeq \alpha \, \tilde{t}^{\beta}.
\]
For $n<4$, it turns out that $\beta=2/n$ and $\alpha^{-n} = (2\beta^2-\beta)$. So for purely matter filled universe with $n=3$, we can estimate the time taken for the universe to expand from $a_{min}$ to any given size $(a)$ to be
  \footnote{The analogous expansion time in GR for the universe to expand to a given size from big bang is $\frac{ 2 a^{3/2}}{3 \sqrt{\Omega_M} } $
 } 
\begin{equation}
 (t-t_{min})   \simeq  \frac{ 0.47  \, a^{3/2}}{\sqrt{\Omega_M}}
\label{expansiontime}
\end{equation}
Hence the time taken for the universe (with just matter) to expand  from the minimum to any given size $(a)$  does not depend on the value of $a_{min}$,  provided $a \gg a_{min}$.  Since the evolution of $a(t)$ is symmetric around $a_{min}$, the above calculation is applicable even for the contraction phase towards the minima. So the total time taken for the universe to contract from a given size to the minima and expand back to the same size will be twice the value given in eq.~\ref{expansiontime}.

For $n=4$ (radiation dominated universe), the above asymptotic solution doesn't converge, so the equations of motion have to be solved numerically with initial conditions provided at the minima. See appendix for elaboration.

\section{Fitting Supernova redshift data}
\label{SUP}

Type Ia supernovae are uniquely qualified to serve as `standard candles' to estimate the distance of the parent galaxy to which they belong. This is because such a supernova explosion is always expected to occur with a fixed peak luminosity as a white dwarf accretes matter from a companion star to reach the Chandrasekhar limit of 1.44 solar masses. Theoretically the absolute magnitude of a Type Ia supernova should be $-19.5$, however there are some variations which can be corrected for by analyzing the shape of the supernova's light curve in the days following the explosion \cite{riess1996precise}. 
The absolute magnitude of a supernova and its apparent magnitude as observed from Earth would together give a distance estimate, which when plotted against  the redshift is known as the Hubble diagram. 

 \begin{figure*}

\includegraphics[width=0.4\textwidth]{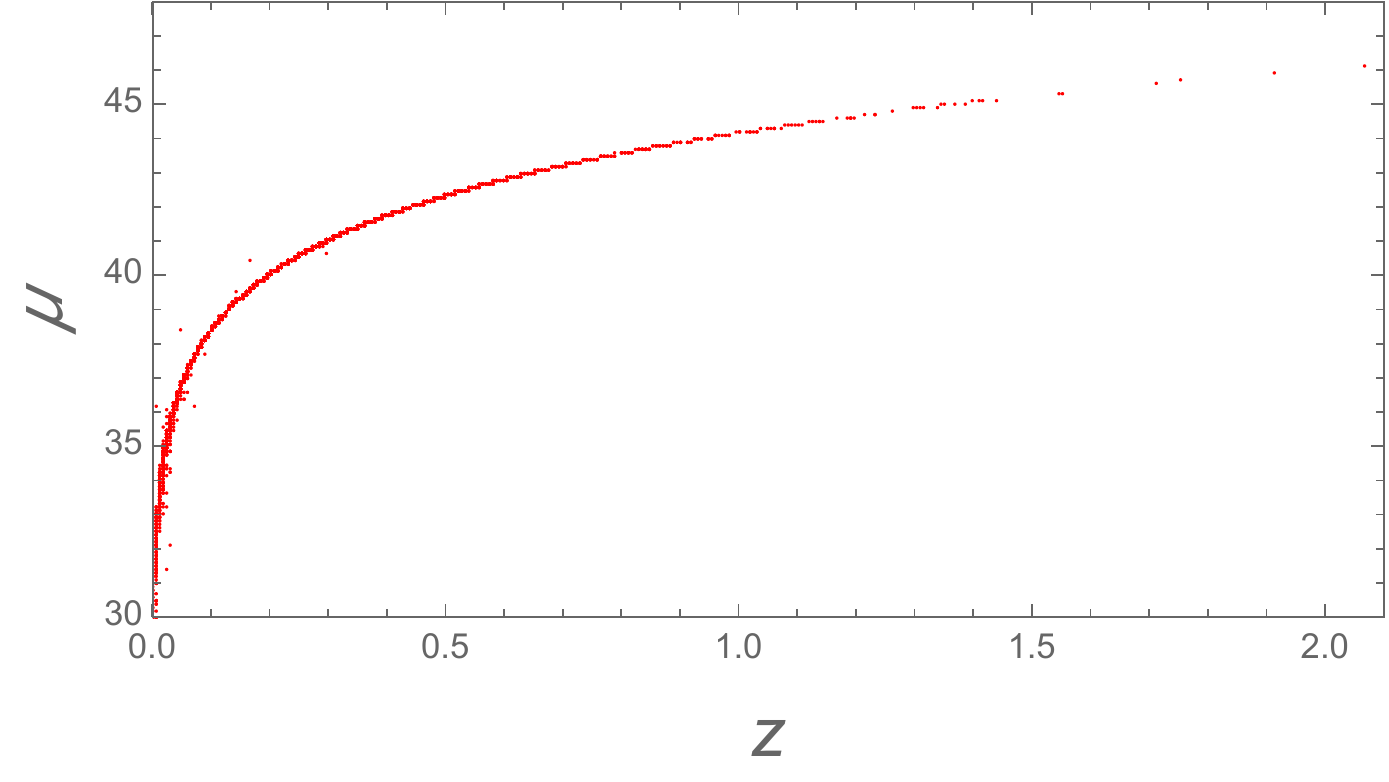} \qquad 
\includegraphics[width=0.4\textwidth]{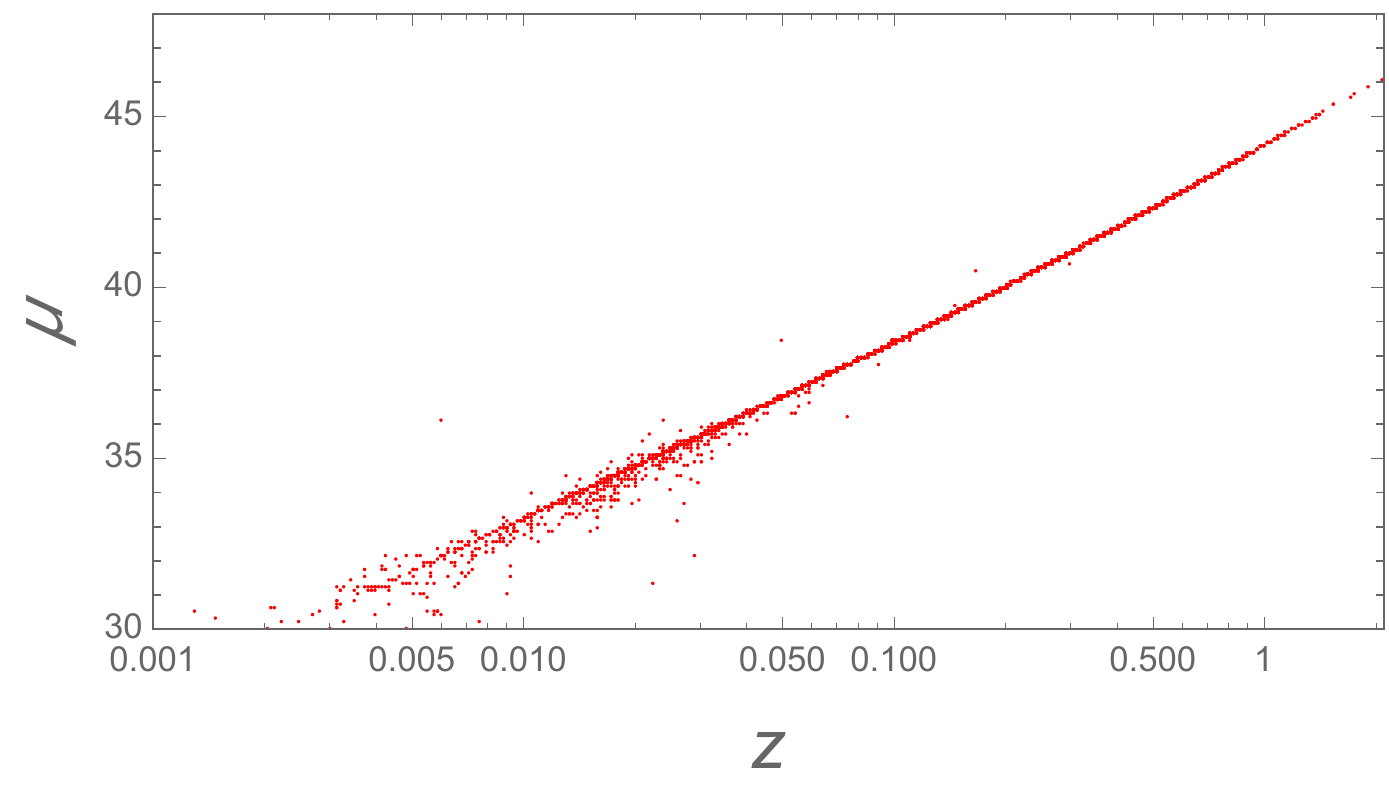}
\\  \qquad \\
\includegraphics[width=0.4\textwidth]{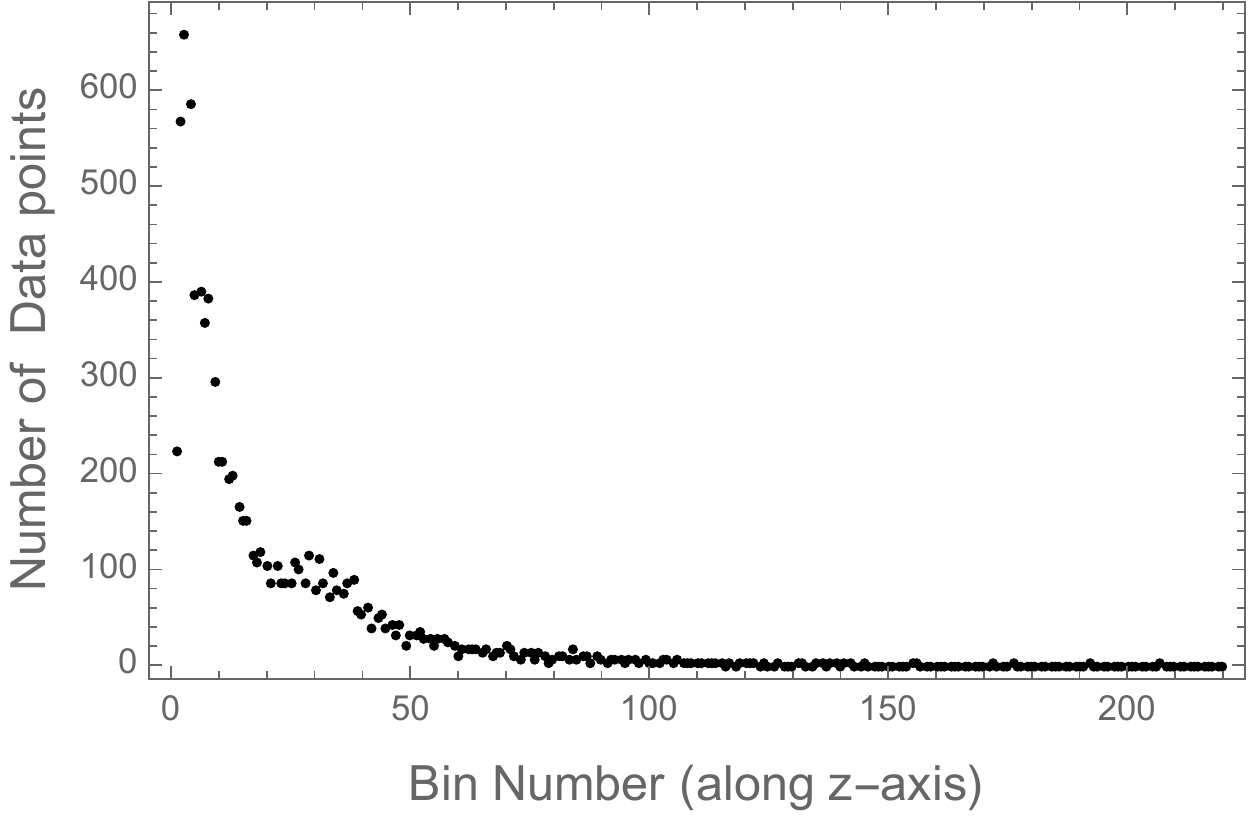} \qquad 
\includegraphics[width=0.4\textwidth]{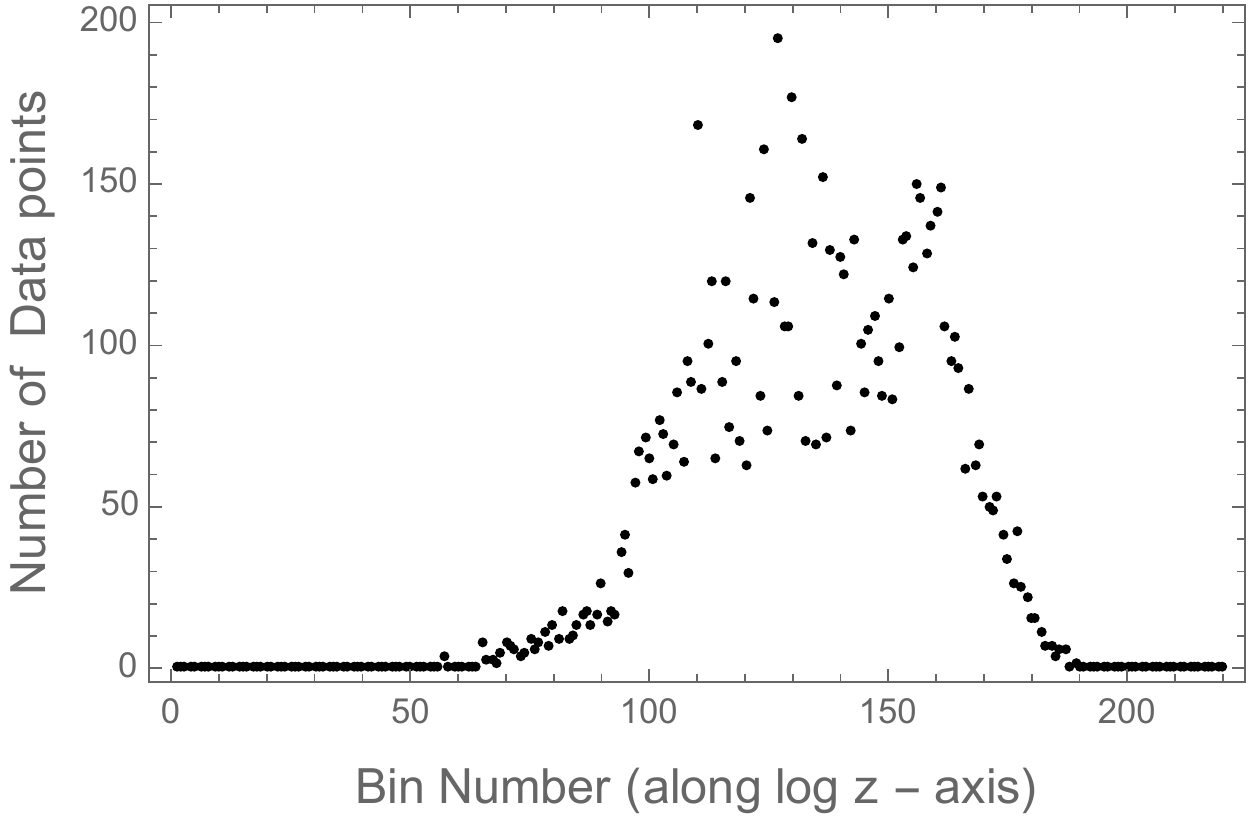}

\caption{Top-left: Hubble diagram of Type Ia supernovae from the Open Supernova catalog containing over 8000 supernovae,  with over 100 of them beyond a redshift of $z=1$. Top-right: The same Hubble diagram is plotted on $\log{z}$ axis.
Bottom-left: Equisized bins are laid out on the $z$-axis of width $\Delta z = 0.01$ ranging from $z=0$ to =2.2 and the number of data points within each bin is plotted. Bottom-right: Equisized bins are laid out on $\log{z}$ axis with a width $\Delta \log{z} =0.05$ in the range $\log{z} \in (-9 , +2)$ to superfluously span the data set, and the number of data points within each bin is plotted. }
\label{Datafig}
\end{figure*}

From the cosmological equations, a Hubble diagram for the model can be generated by following these steps:

1. Numerically integrate the cosmological equation with given initial conditions  at $t=0$ to obtain $a(t)$ for all $t$ beyond the most recent minima. 

2. The wavelength of a photon traveling through the universe linearly scales up with the scale factor $a(t)$, and the red shift  $z$ of the light emitted by a supernova at some time $t$ in the past as observed at $t=0$ is   
\[1+z= 1/a(t)\]

3. The comoving distance travelled by light  from a supernova  at time $t$ to Earth at $t=0$  is
\begin{equation}
\chi = \int_t^0 \frac{dt'}{a(t')} =  \int_0^r \frac{dr'}{\sqrt{1+ \Omega_k r'^2}}
\nonumber
\end{equation}
and the radial coordinate $r$ of that supernova is $S(\chi)= \{ \chi,\, |\Omega_k|^{-1/2} \sin (\sqrt{|\Omega_k|} \chi ), \, |\Omega_k|^{-1/2} \sinh ( \sqrt{|\Omega_k|} \chi ) \} $ for $\Omega_k=\{ 0,-,+\}$ respectively.

4. The luminosity distance $d_L$ is defined in terms of the source luminosity $L$ and the observed energy flux $F$ such that  $4 \pi d_L^2  F = L$, which can be computed for the expanding universe in terms of the redshift to be
\[d_L =  S(\chi) (1+z) \]

5. The magnitude modulus of a supernova is defined as $\mu \equiv $ absolute magnitude - apparent magnitude. From the definition of luminosity magnitude, the relationship between the magnitude modulus and the luminosity distance is given by
\begin{equation}
 \mu = 5 \log _{10}( d_L) + 25  
 \label{mu}
 \end{equation}
 where $d_L$ is measured in Megaparsecs (Mpc). \footnote{The Hubble constant is generally measured in units of km/s/Mpc. The $d_L$ obtained from the model should be multiplied by a factor $c H_o^{-1}$ to make it physically dimension-full.}
 
 To generate the Hubble diagram for the model, we need to first back-calculate the value of $t$ for any given $z$ from the numerical solution of $a(t)$, and then evaluate $\chi$, $d_L$ and $\mu$ from the above steps to plot $\mu$ vs $z$.
    
Plotting the Hubble diagram of the observed data is more involved, because the observed magnitude of supernova has to be corrected for absorption effects of interstellar gas on a case to case basis. We shall however not be concerned of those details, and just use the reported post-processed magnitude from a large data set. The open supernova catalog  \cite{SupernovaDataset} (https://sne.space/) contains data from over 8000 Type Ia supernovae providing the details of each individual measurement and an estimate of $d_L$ after such corrections are applied. The magnitude modulus $\mu$ is then calculated from eq.~\ref{mu} for each supernova. The list of supernovae names along with their redshift and magnitude modulus is given in the supplementary information, and the associated Hubble diagram is plotted in the top left panel of fig.~\ref{Datafig}.

\begin{figure*}
\textbf{Model accuracy for the standard model $\LCDM$ parameters} \\
\qquad \\
\includegraphics[width=0.32\textwidth]{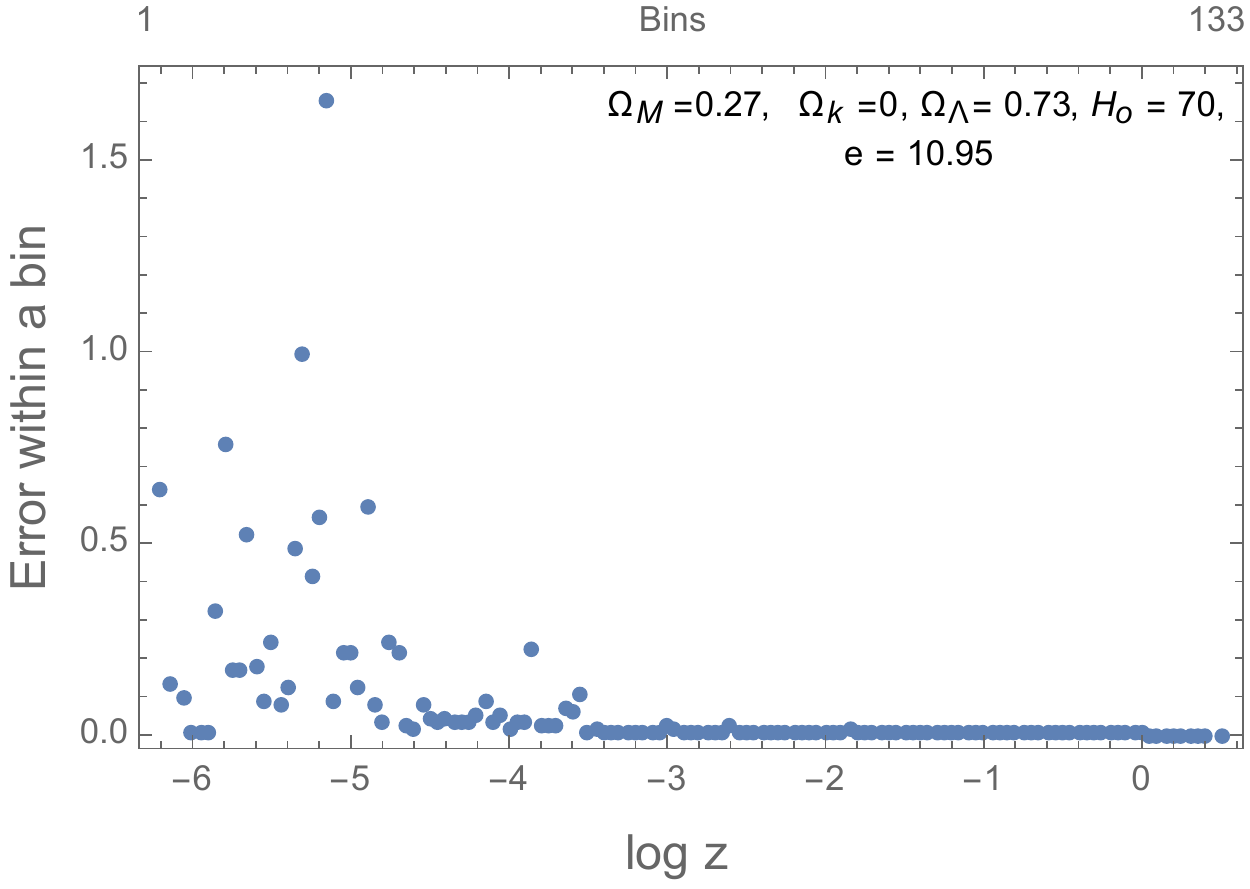} 
\includegraphics[width=0.32\textwidth]{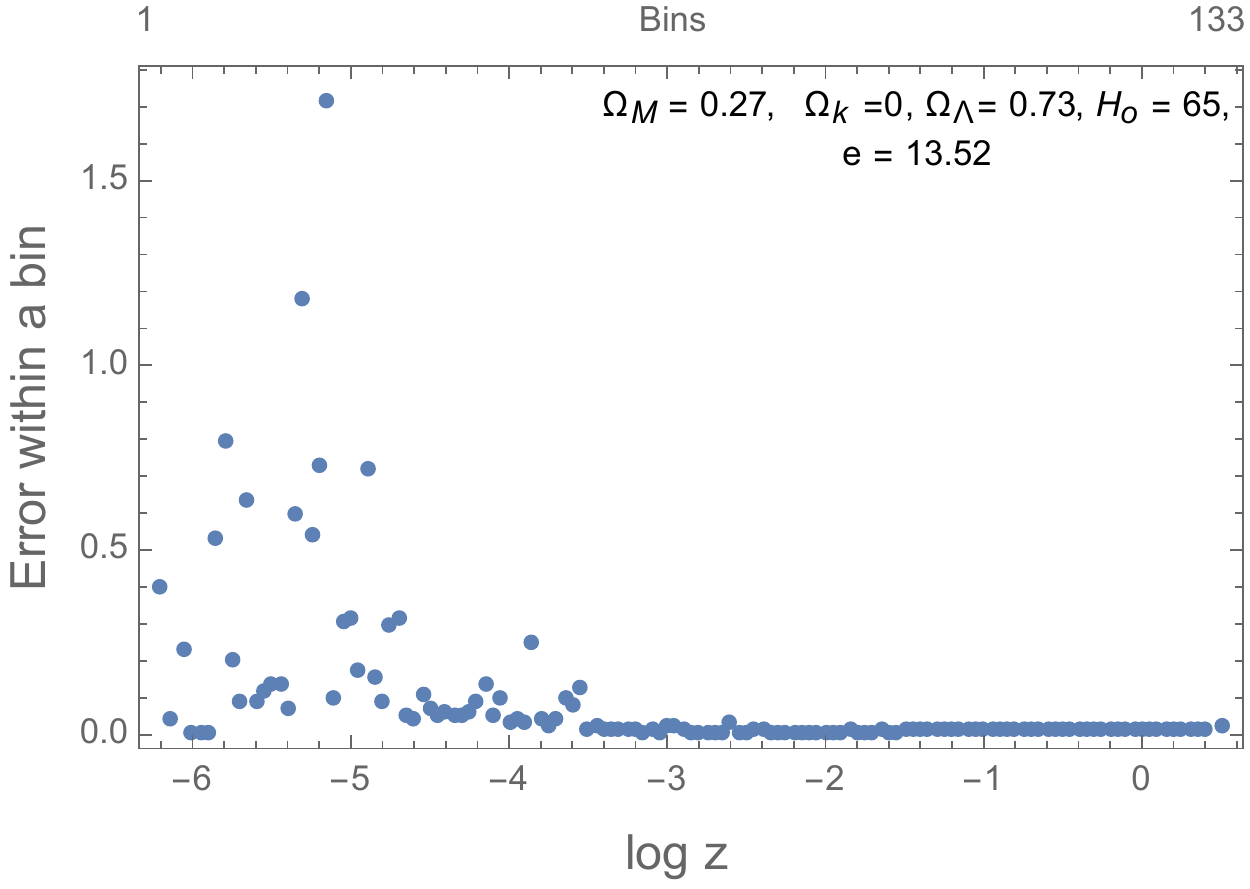}
\includegraphics[width=0.32\textwidth]{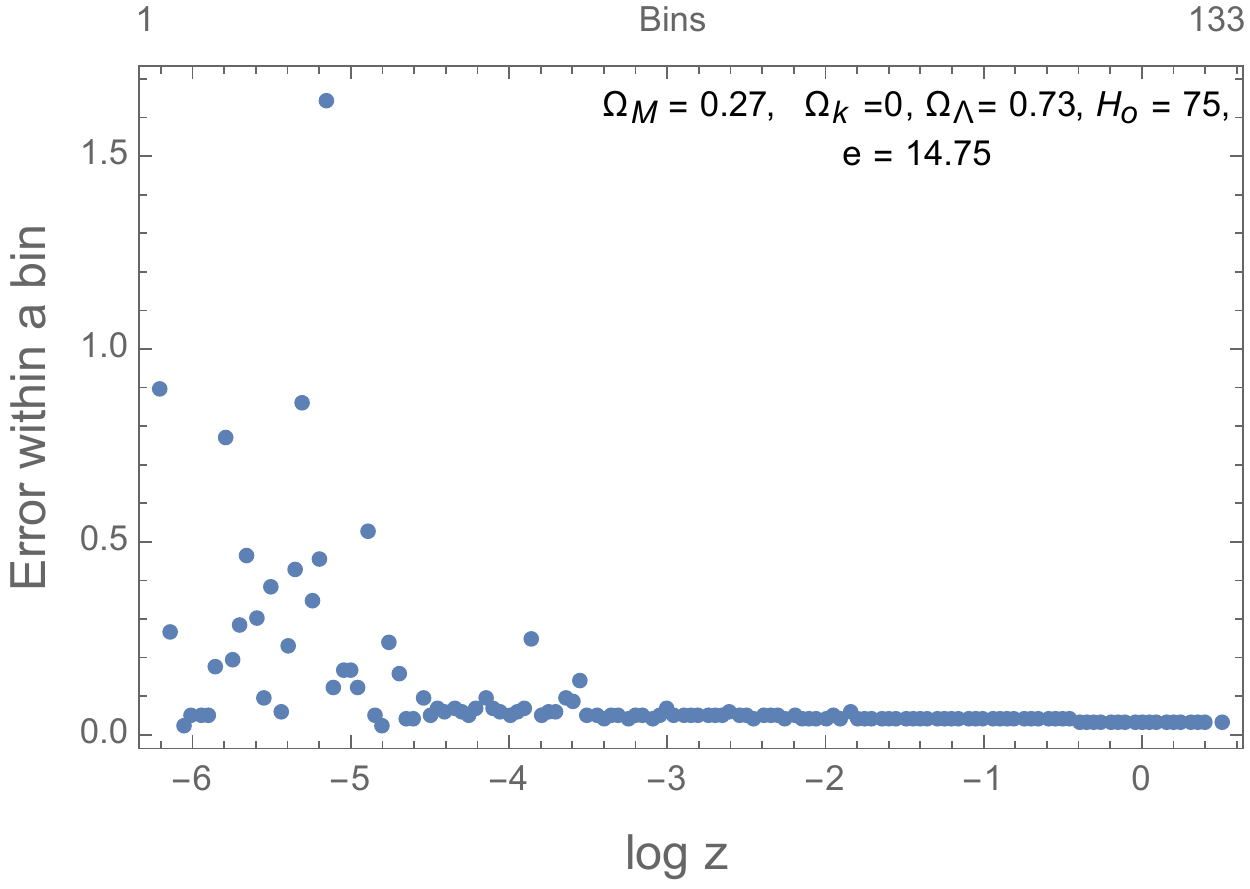}
\\  \qquad \\
\includegraphics[width=0.32\textwidth]{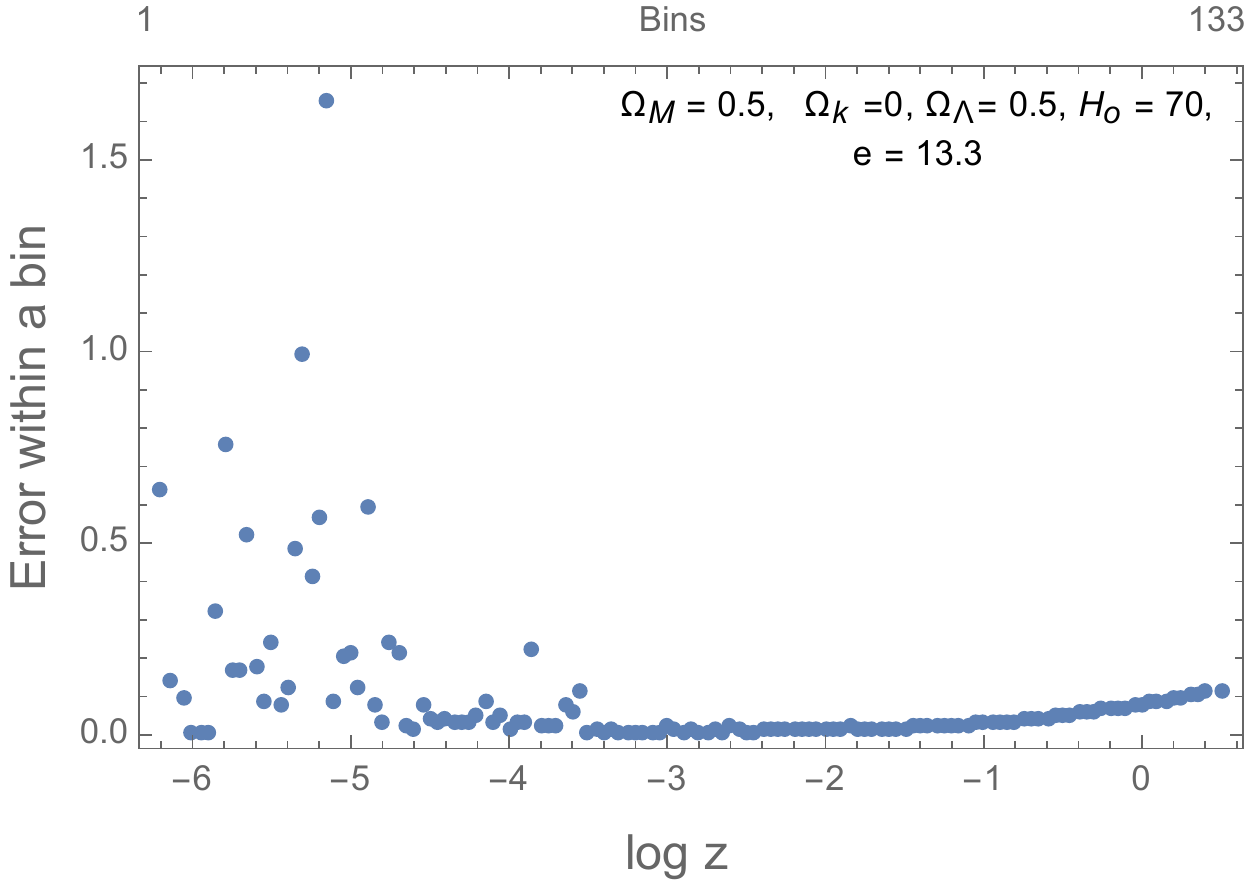} 
\includegraphics[width=0.32\textwidth]{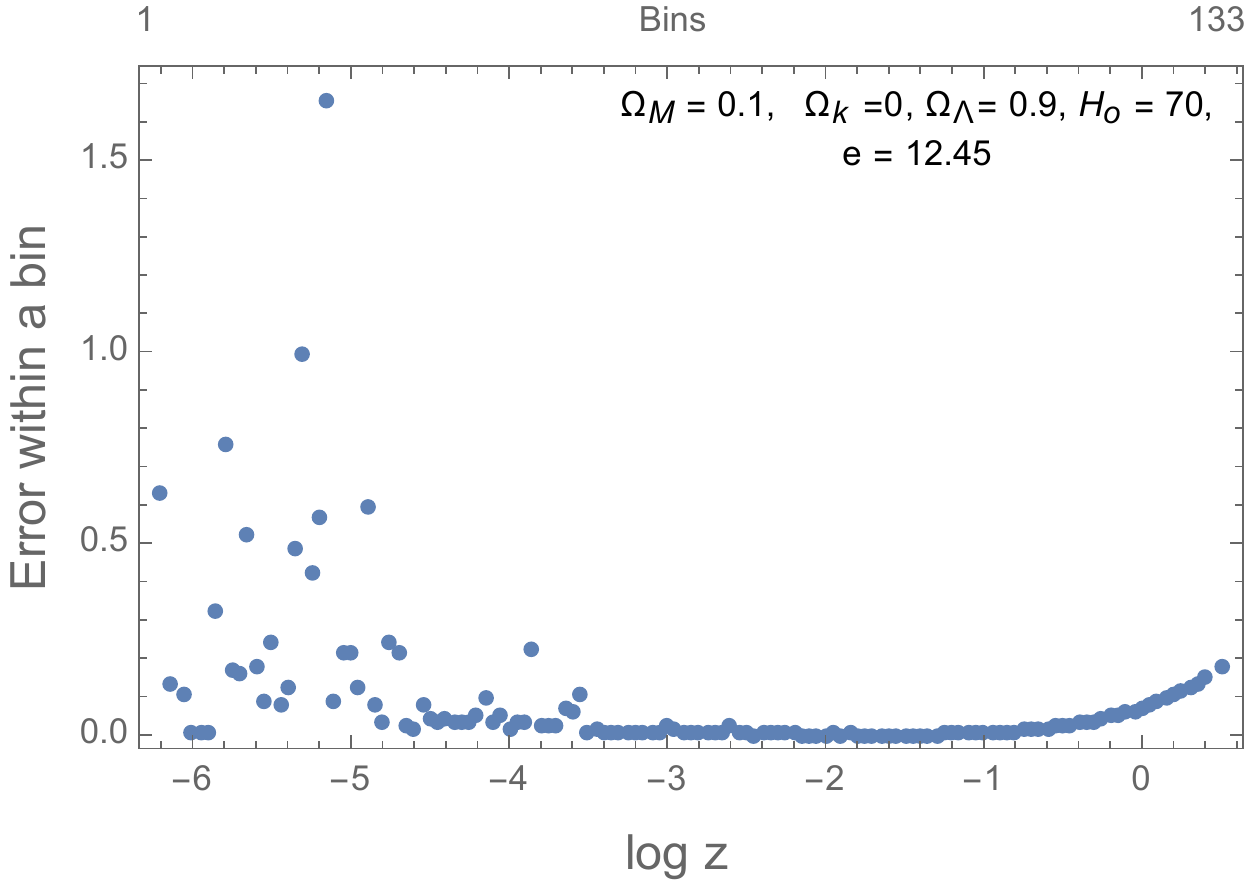}
\includegraphics[width=0.32\textwidth]{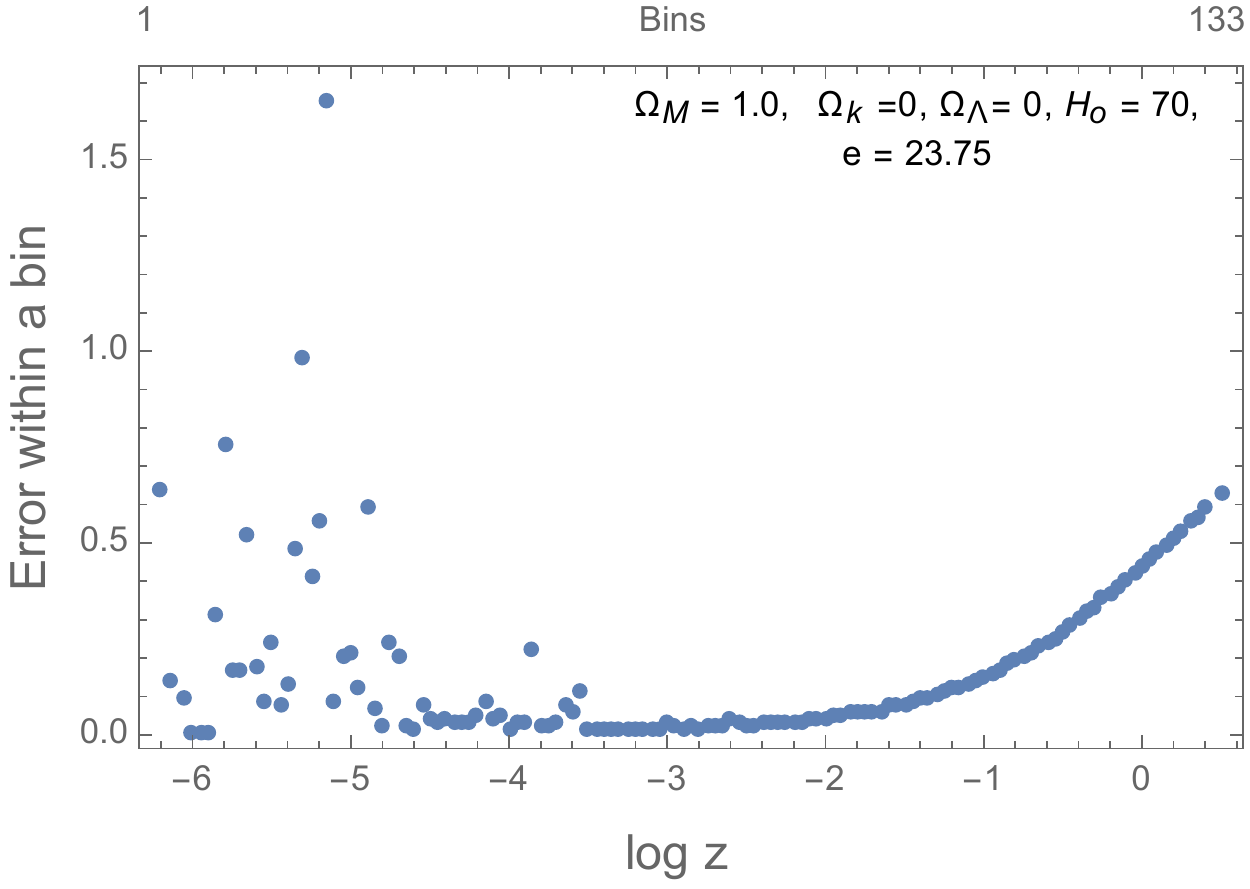}
\\ \quad \\
\includegraphics[width=0.32\textwidth]{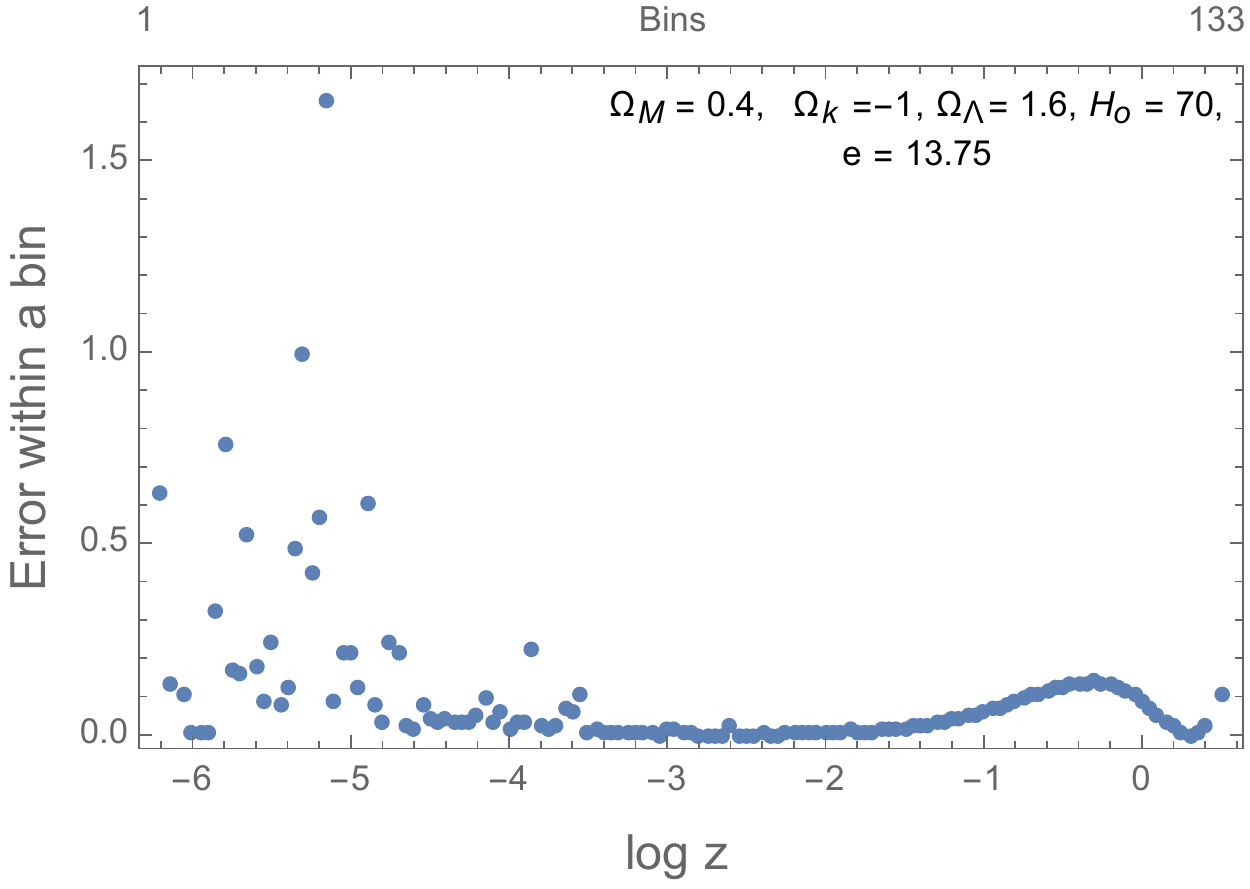} 
\includegraphics[width=0.32\textwidth]{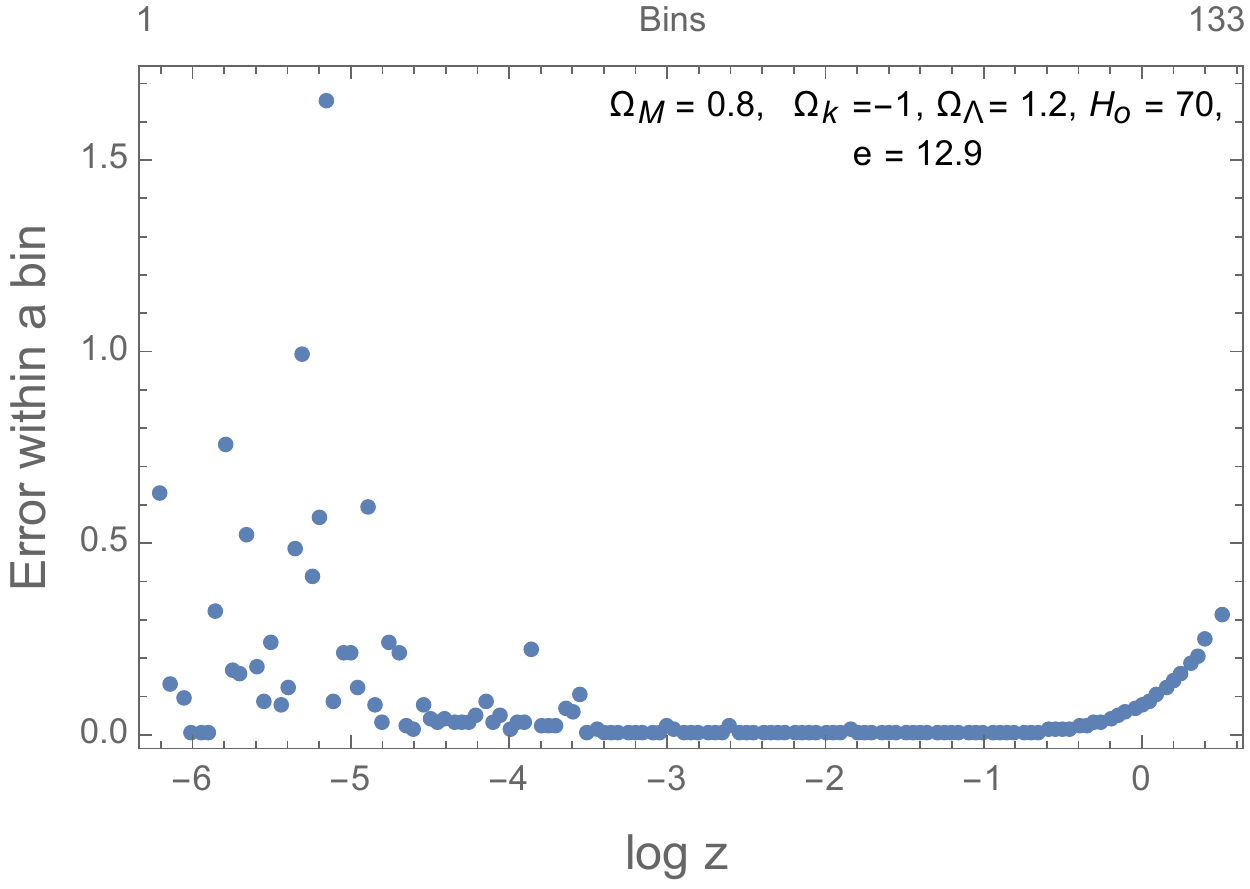}
\includegraphics[width=0.32\textwidth]{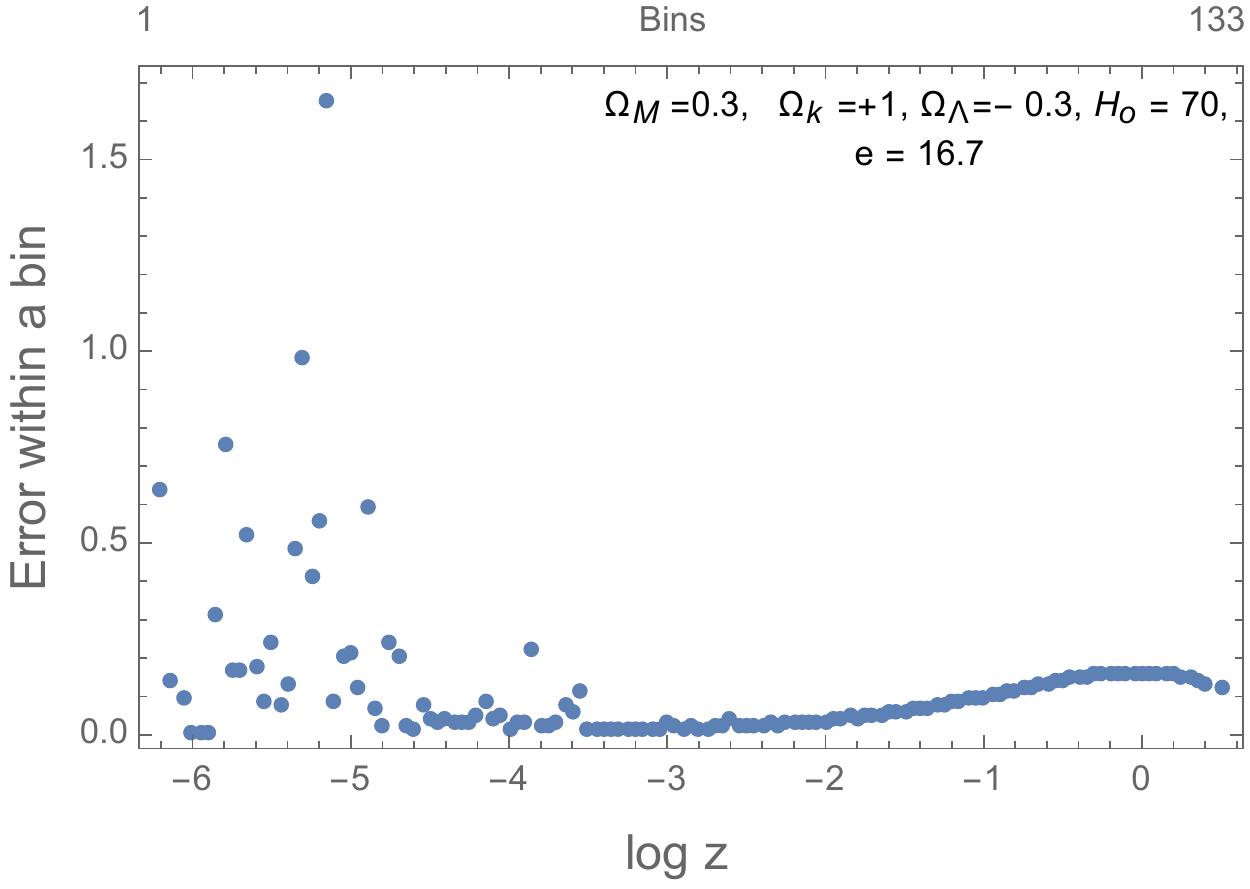}
\\ \quad \\
\caption{The model-data discrepancy $ \big \langle |\mu_{model} -\mu_{data}|^2  \big\rangle_{\textrm{bin}} $ at various scales of the universe (quantified as bins)  is plotted for the standard model with some illustrative  parameter values. The bins are laid out on $\log z$ axis with a width of $\Delta \log z =0.05$ and the data is distributed across the bins as  shown in the bottom-left panel of fig.~\ref{Datafig}. The best fit parameters are   $\Omega_M=0.27$, $\Omega_k=0$, $\Omega_{\Lambda}=0.73$, $H_o=70$ km/s/Mpc, which closely match the consensus estimates.  } 
\label{FRWer}
\end{figure*}

\begin{figure*}
\textbf{Model accuracy for THED parameters} \\ 
\qquad \\
Oscillating (near Green Surface) \qquad  \qquad \qquad Ever Expanding (Best fit) \qquad   \qquad \qquad \qquad Moderate fit \qquad \qquad \qquad \qquad   \\  

\includegraphics[width=0.32\textwidth]{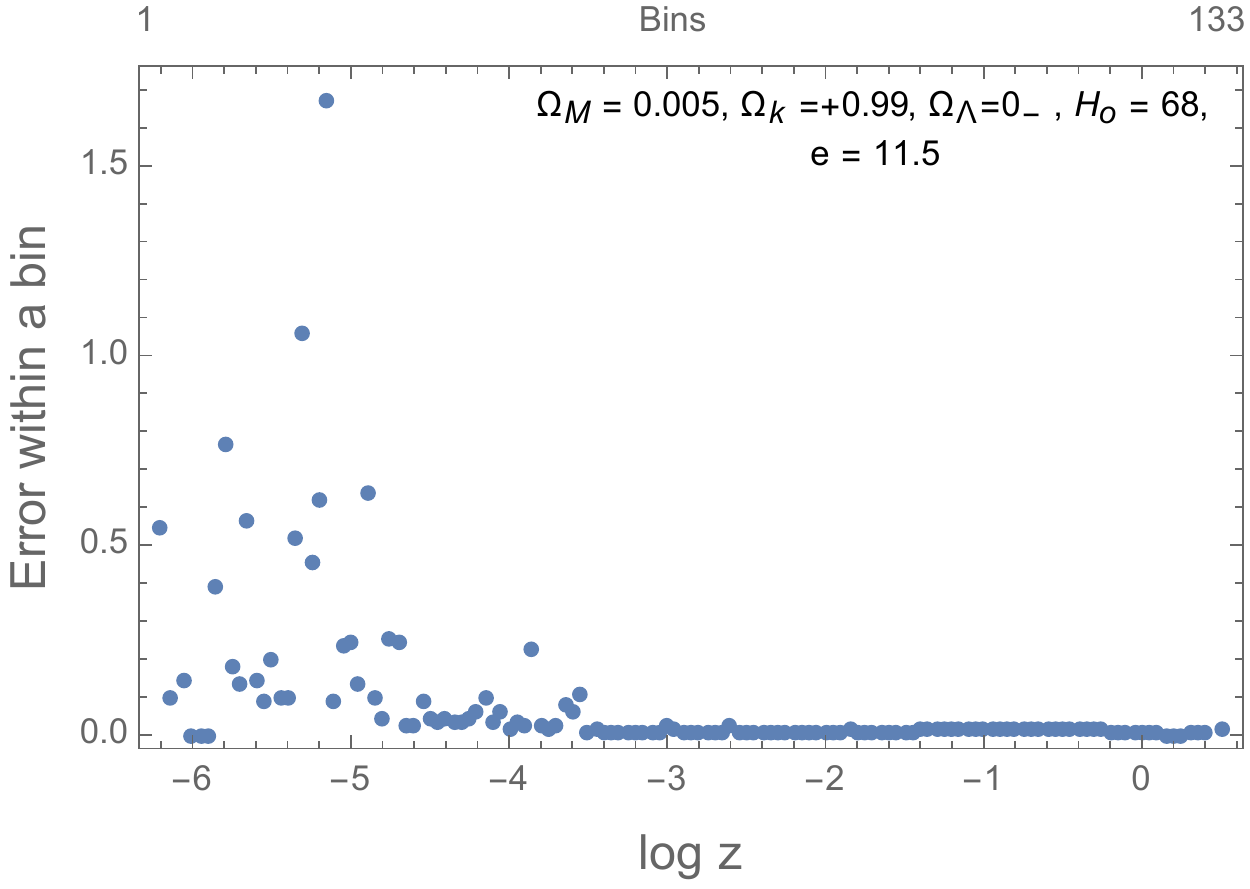} 
\includegraphics[width=0.32\textwidth]{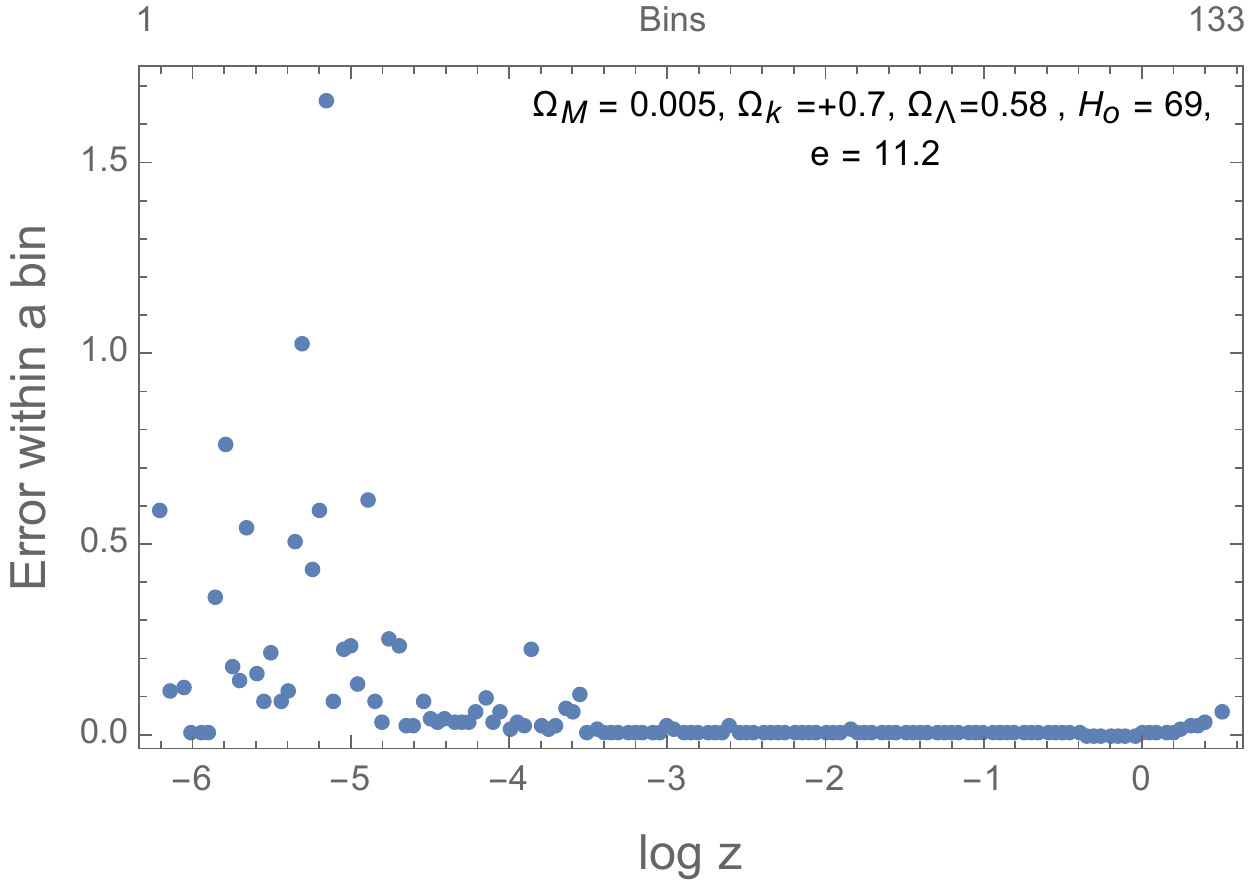}
\includegraphics[width=0.32\textwidth]{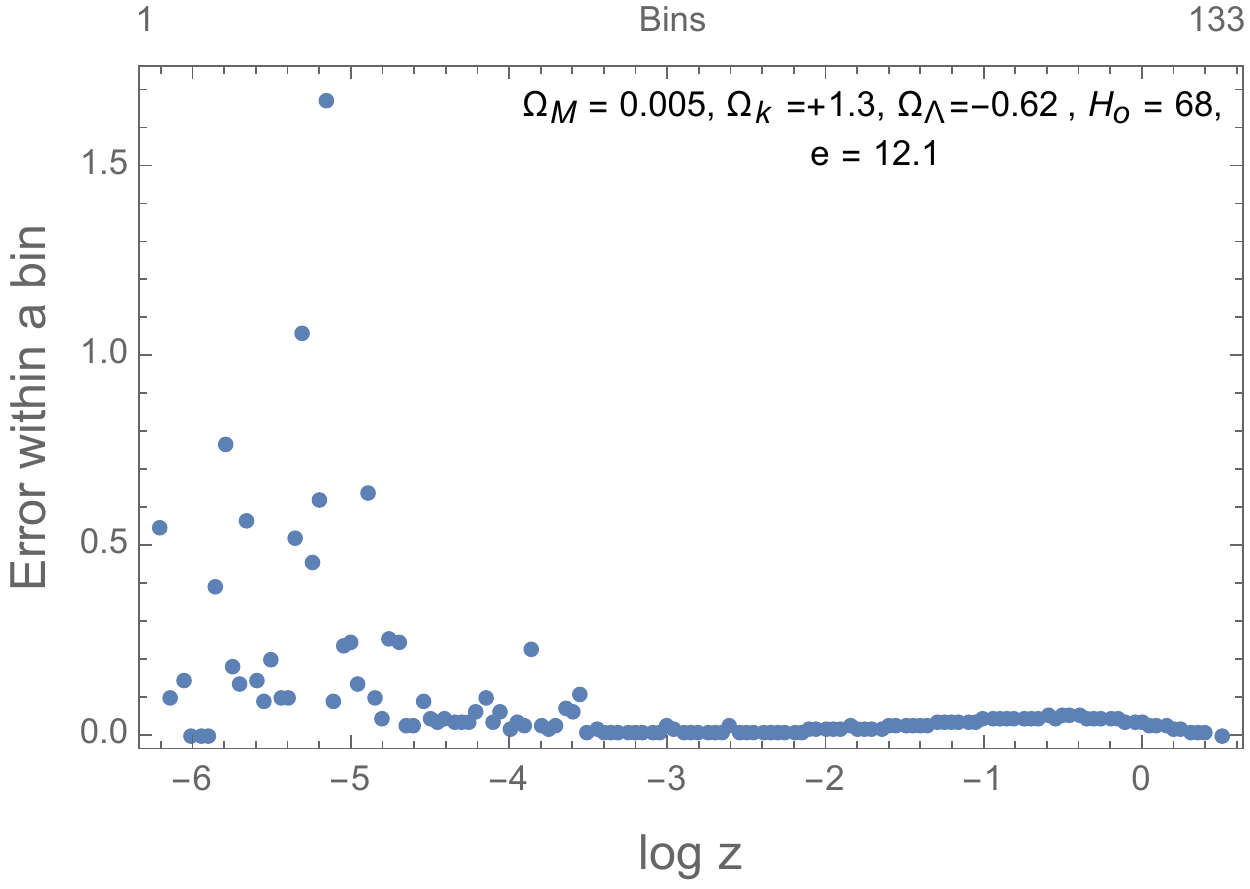}
\\  \qquad \\
\includegraphics[width=0.32\textwidth]{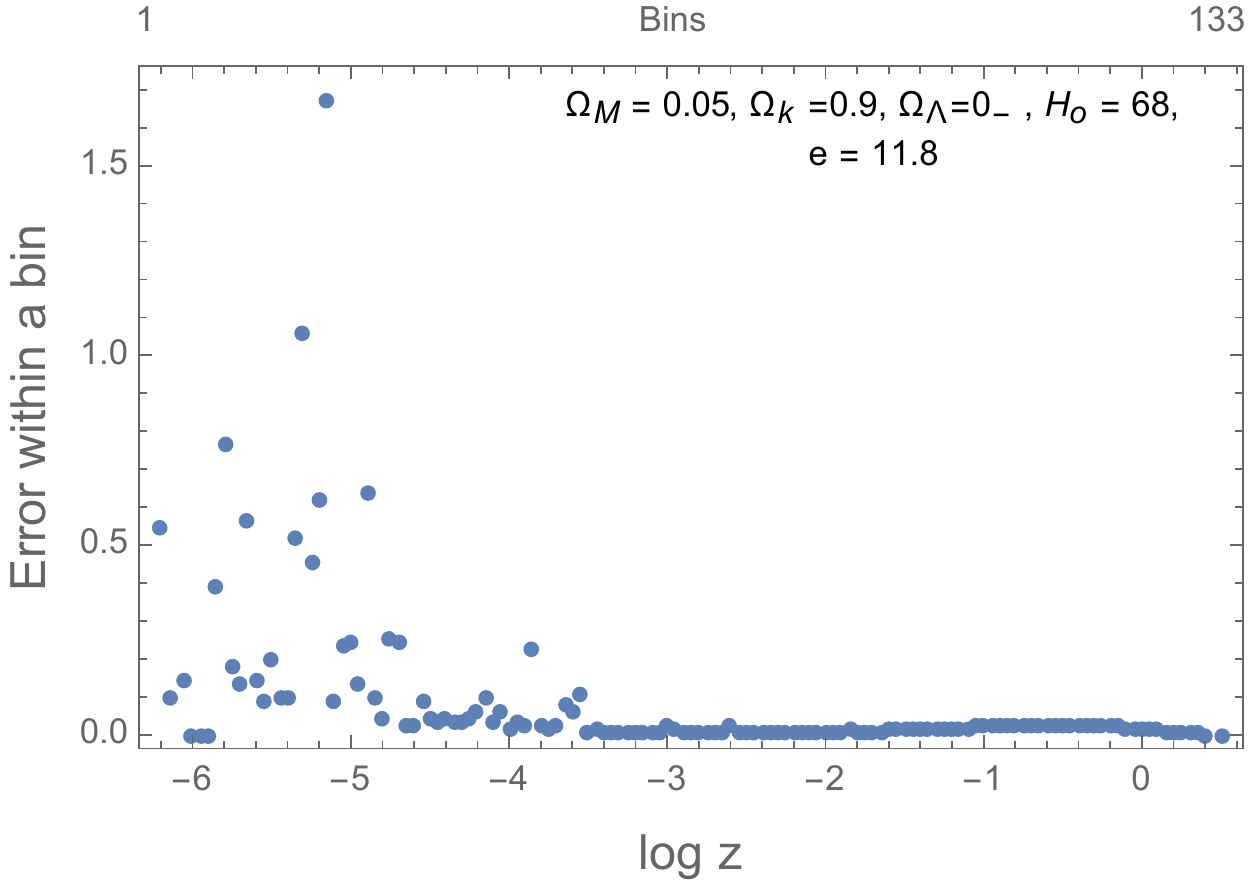} 
\includegraphics[width=0.32\textwidth]{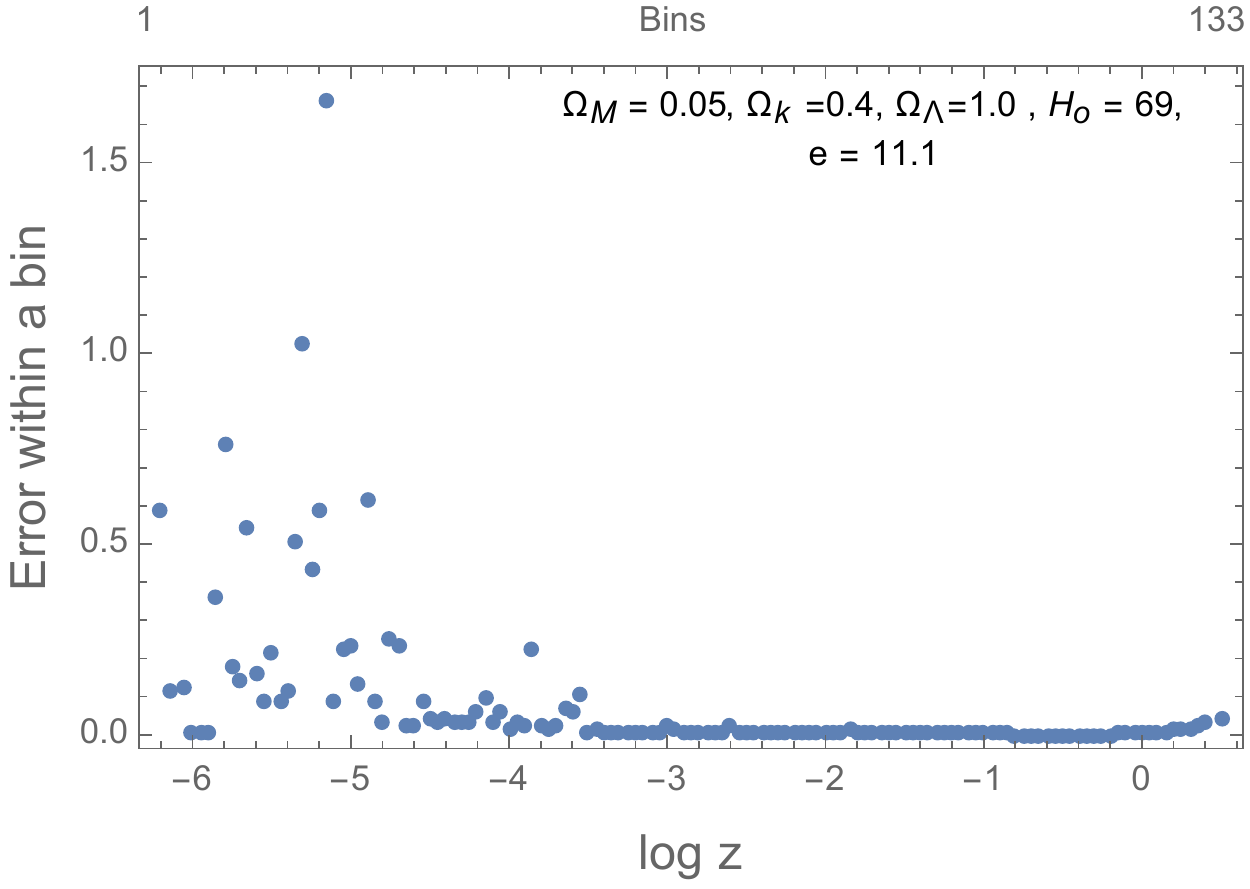}
\includegraphics[width=0.32\textwidth]{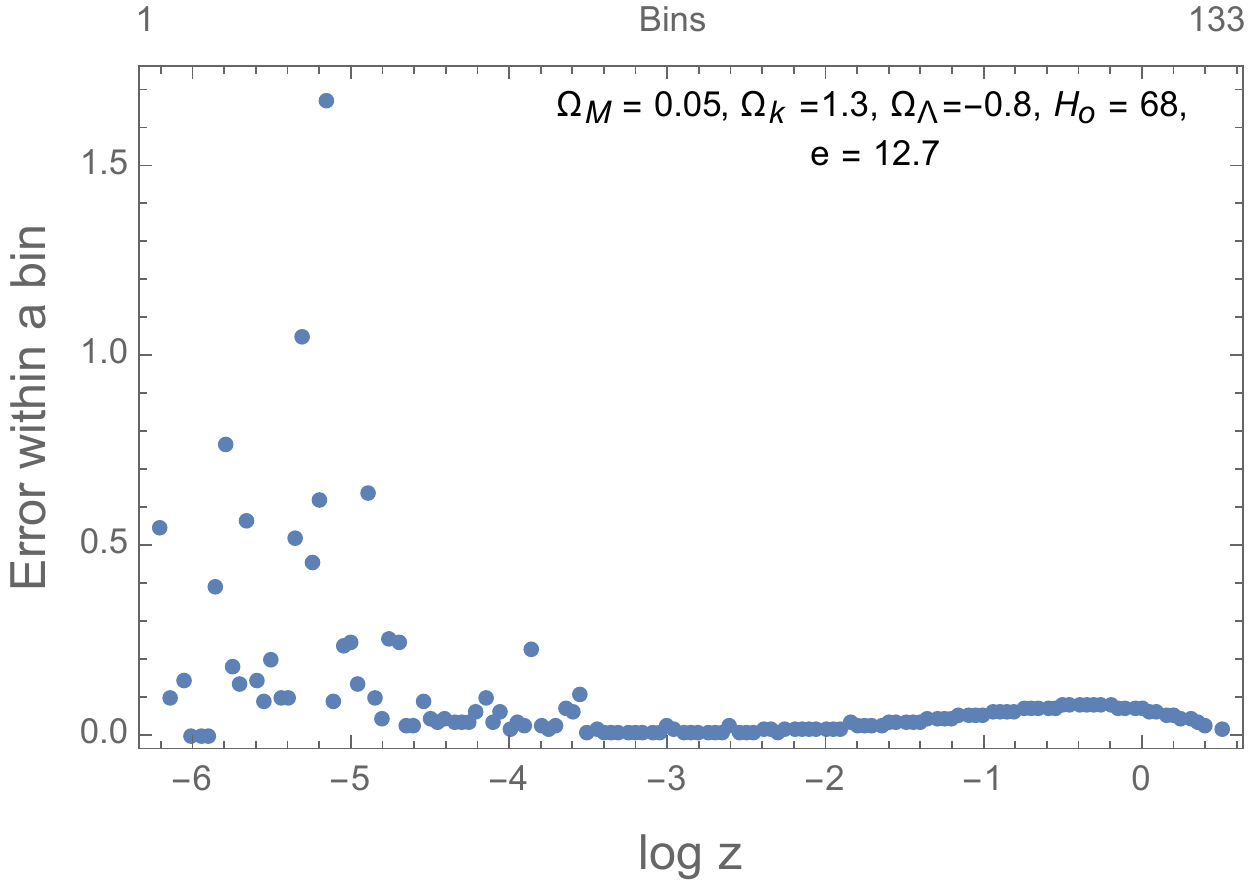}
\\ \quad \\
\includegraphics[width=0.32\textwidth]{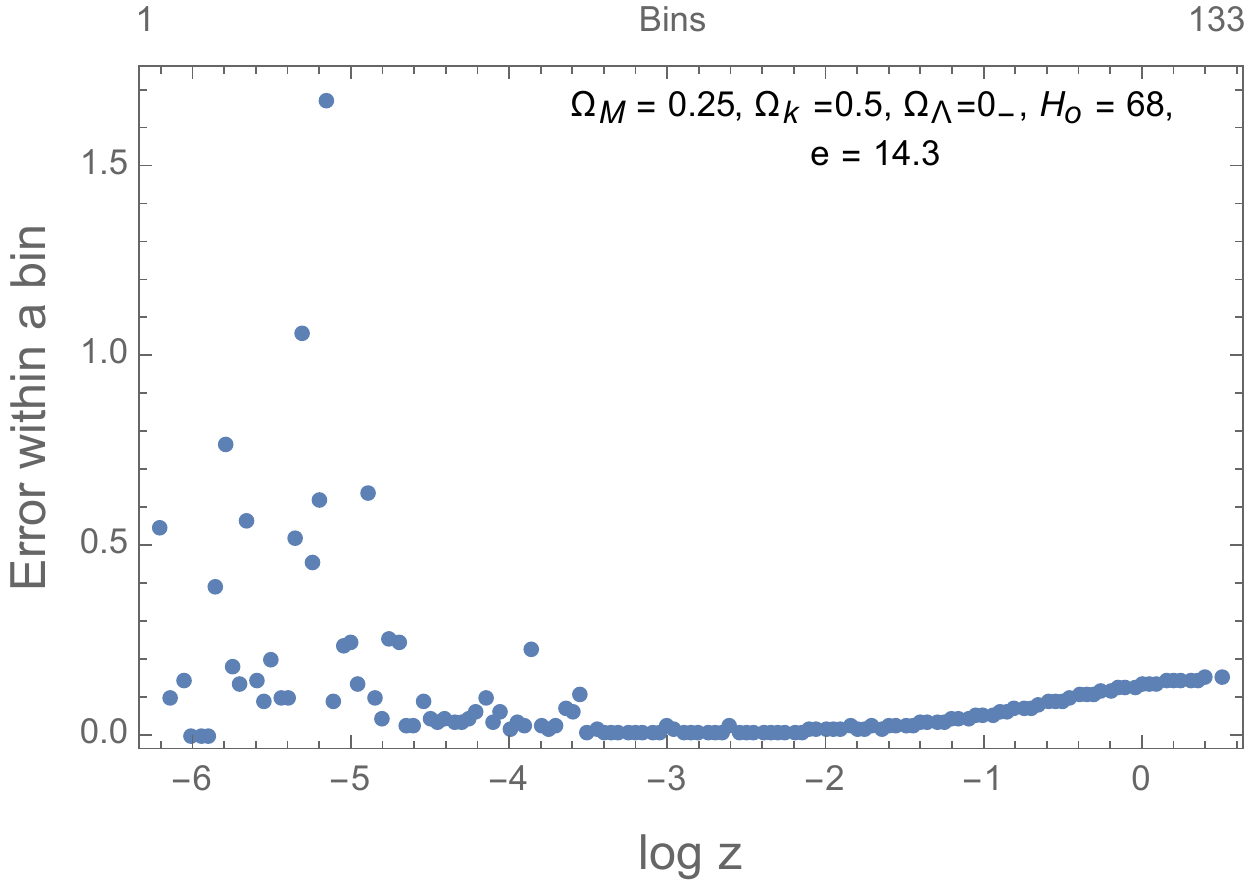} 
\includegraphics[width=0.32\textwidth]{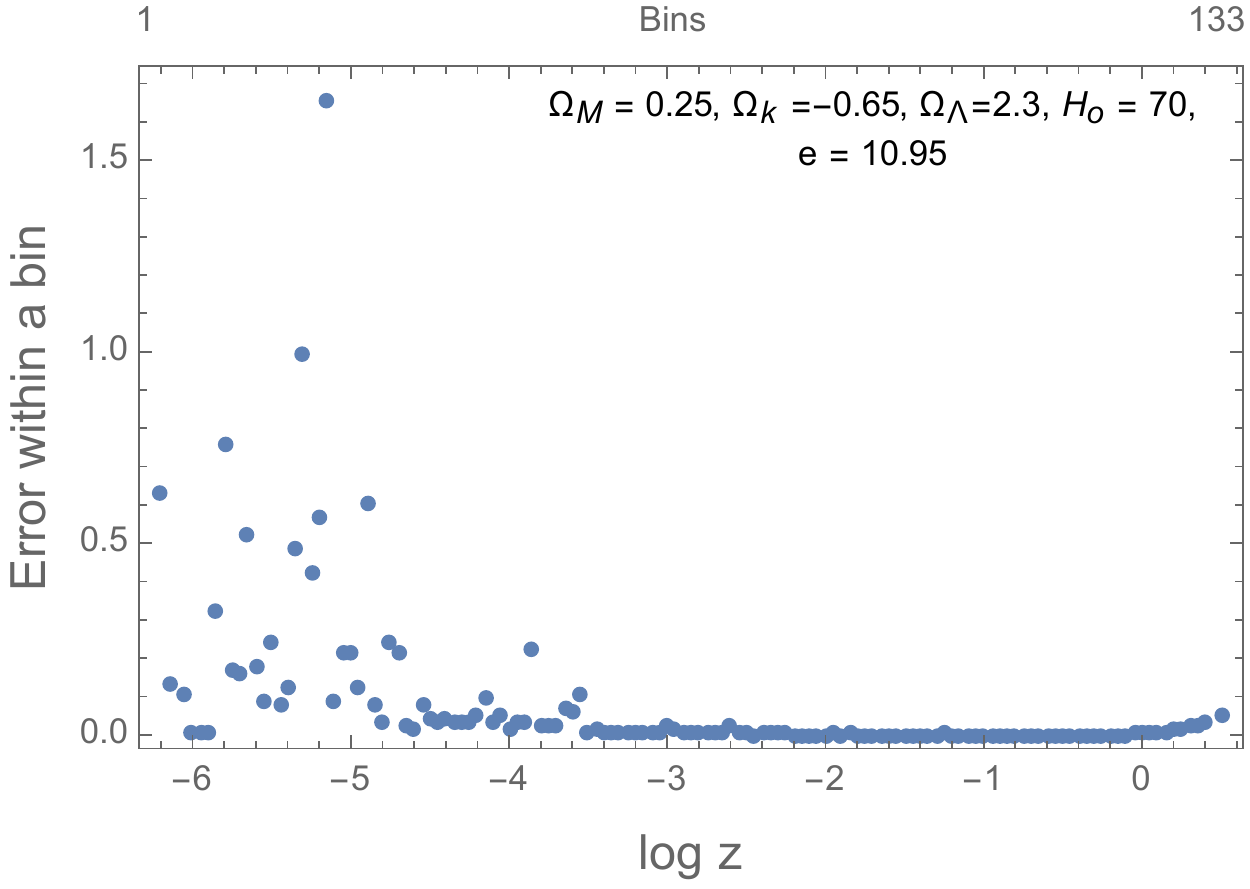}
\includegraphics[width=0.32\textwidth]{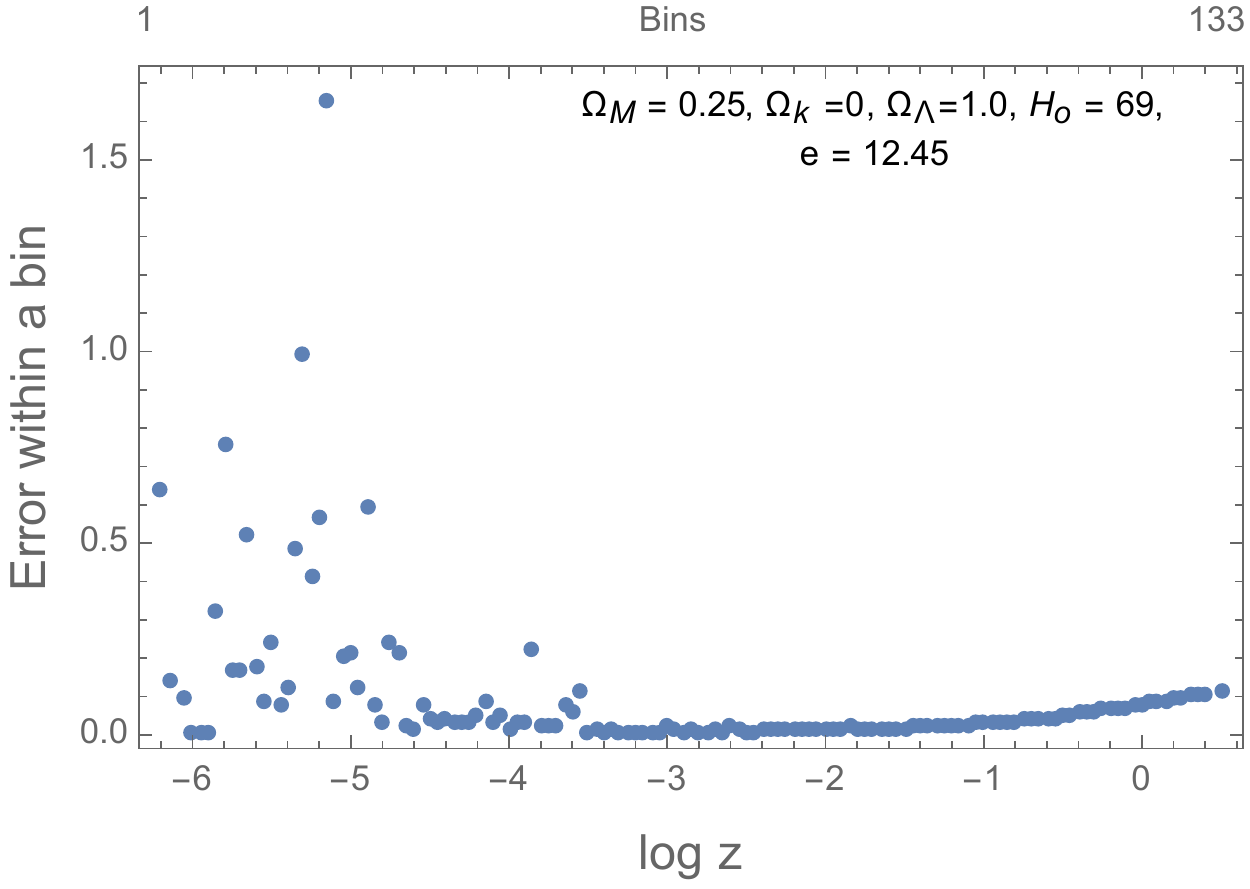}
\\ \quad \\
\caption{The model-data discrepancy   $ \big \langle |\mu_{model} -\mu_{data}|^2  \big\rangle_{\textrm{bin}} $ at various scales of the universe (quantified as bins)  is plotted for THED gravity with some illustrative  parameter values. The bins are laid out on $\log z$ axis with a width of $\Delta \log z =0.05$ and the data is distributed across the bins as  shown in the bottom-left panel of fig.~\ref{Datafig}. In the left column, the parameters are very close to the green surface with $\Omega_{\Lambda}$ being a very small negative number, and these correspond to oscillatory solutions with very large periodictiy. The middle column shows the parameters that best fit the data for chosen values of $\Omega_M$, but these correspond to ever expanding solutions because these parameters lie above the green surface in fig.~\ref{paramfigure}. }    
\label{ThedEr}
\end{figure*}

\subsection{Binned analysis on log $z$ - axis}

The Supernova dataset has been steadily growing to include larger redshifts from more distant supernovae, and the parameters of the standard model of cosmology  $\LCDM$ has been fitted with increasing precision on this dataset \cite{riess2004type}. Although the consensus opinion favors a universe dominated by dark energy with an accelerated expansion \cite{riess2004type}, modified analyses reveals that the Supernova data (when considered in solo) is also consistent with a non-accelerating universe with zero dark energy \cite{nielsen2016marginal}. Since we now have access to a very large data set containing more than 8000 data points with over 100 data points with $z>1$  \cite{SupernovaDataset}, we can  examine the model-accuracy corresponding to different scales of the universe separately. 
   
To estimate the model parameters that best fit the data, the standard strategy is to take into account the uncertainty in the measurement of $z$ and $\mu$ for each data point, and compute the parameters that maximizes the likelihood of the dataset. But here we shall take a simpler approach by ignoring the error-bars in the data, and segregating the data into appropriately sized bins with sufficiently many data points within each bin. The range of  $z$-values within each bin gives the scale of the universe represented by that bin.  Any model-data discrepancy within a bin is a measure of the deviation of the model from the data at that scale of the universe. The model-data discrepancies from all bins should be equally weighted to obtain a measure of overall model accuracy. Just because more data is available at some scale, that specific scale should not over-represented in evaluating the model accuracy.

 \begin{figure*}
\includegraphics[width=0.8\textwidth]{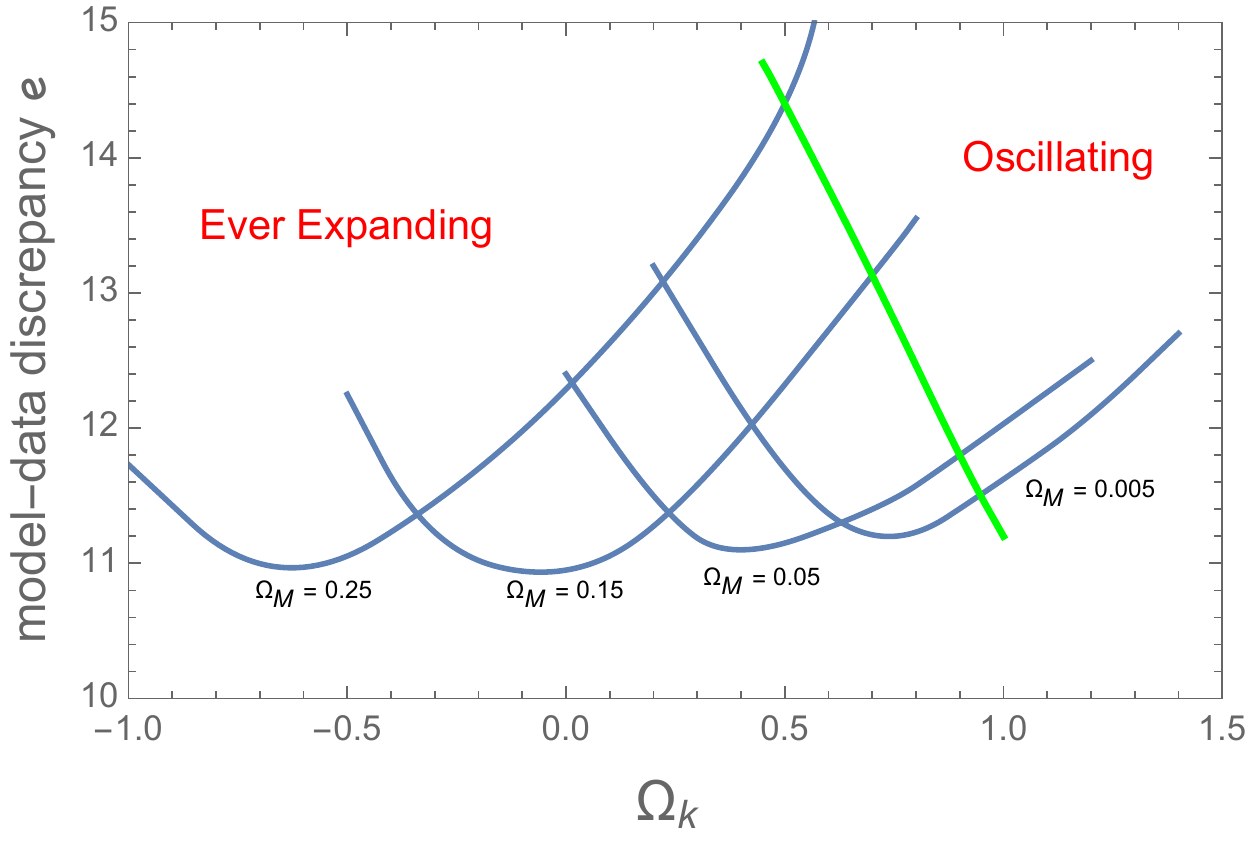} 
\caption{The model-data discrepancy is plotted for four different values of $\Omega_M$ as a function of $\Omega_k$. The value of $\Omega_{\Lambda}$ is chosen so that the parameters fall on the red surface in fig.~\ref{paramfigure} with $a_{min}\rightarrow 0$. The green line marks the position where $\Omega_{\Lambda} =0$. The oscillating solutions lie to the right of the green line and to its left lie all solutions that expand for ever from the bounce at $a_{min}$.  }
\label{ThedFits}
\end{figure*}

The binning-choice essentially determines the way we evaluate the model. We could simply choose bins of constant width along the $z$-axis, but note that the number of data points in the bins with low $z$ values is much higher  than those with large $z$ (bottom-left panel of figure \ref{Datafig}). Alternatively, we  could choose bins of variable sizes along the $z$ axis such that the each bin has a fixed number of data points. We should however not segregate the scales of the universe based on the existing number of data points at various scales; rather the segregation of scales (or bins) should be based on some theoretically appealing feature of the data. 

By definition, the difference in \emph{magnitude} of two light sources is explicitly a measure of the perceptual difference in brightness between them. So any uncertainty in the measurement of magnitude can be expected to be a constant (or at least uncorrelated to the magnitude itself)--this is indeed well established in psychophysics as the Weber-Fechner law \cite{fechner1948elements}.  This suggests a natural way to segregate the data--into equisized  bins along the measured luminosity magnitude, so that the uncertainty in all the bins will be the same. However, since the magnitude modulus $\mu$ is  dependent on the redshift $z$, we can rescale the $z$-axis so that $\mu$ appears linear along the rescaled $z$-axis, and the data can be binned along the rescaled $z$-axis. The Hubble diagram from the dataset is plotted in the top-left panel of figure \ref{Datafig} and the top-right panel shows the same Hubble diagram on a logarithmic-$z$-axis.  Notice that on a logarithmic-$z$-axis, the dependent variable $\mu$ is almost linear.  So, the natural binning-choice is to segregate the data into equisized bins along  $\log{z}$ axis.

Every bin is considered equal in evaluating the model accuracy irrespective of the number of data points it contains (as long as there exists at least one data point).  The error in $\mu$ for the $i$-th data point $\{z_i, \mu_i \}$ is  computed as $|\mu_{model} (z_i) - \mu_i |^2$, which is then averaged within  each  bin, and finally aggregated over all the bins to obtain a measure of model-data discrepancy $e$.    

\begin{equation}
e= \sum_{\textrm{bin}}  \big \langle |\mu_{model} -\mu_{data}|^2  \big\rangle_{\textrm{bin}} 
\label{modelaccuracy}
\end{equation}

We could view the above described  measure of model-data discrepancy with skepticism because the data within each bin is significantly scattered, especially in the very low $z$ range (see top-right panel of fig.~\ref{Datafig}). Nevertheless, since the intrinsic variability in the data within each bin would affect all models (irrespective of the choice of model parameters) in the same way, we do not expect it to have significant effect on estimating the best fit model parameters.

To gain confidence in  eq.~\ref{modelaccuracy} as a legitimate measure of model accuracy, we shall test it out by estimating the best fit parameters of the standard model. Let us choose the bins to be laid out on the $\log z$ axis with a width of $\Delta \log z =0.05$ as shown in bottom-left panel of fig.~\ref{Datafig}. However, due to sparsity of data at the extremes, we can only populate a total of 133 bins and the model accuracy is evaluated only in the range of $z \in (0.001, 1.7)$.  Figure ~\ref{FRWer} shows the error within each bin $\big \langle |\mu_{model} -\mu_{data}|^2  \big\rangle_{\textrm{bin}}$ for different sets of parameters in $\LCDM$. The minimal value of $e$ is attained for $\Omega_\Lambda=0.73$, $\Omega_M=0.27$, $\Omega_k=0$, $H_o=70$ km/s/Mpc conforming with the standard consensus. This validates eq.~\ref{modelaccuracy} as a measure of model accuracy and can hence be trusted to estimate parameters in THED.

\subsection{Parameter Estimation}

To estimate the parameters in THED gravity that would fit the supernovae data, let us keep note of the following. (i) The parameters are restricted to lie very close to the red surface in fig.~\ref{paramfigure} (very small $a_{min}$) so that the universe was extremely hot and dense in the past. This implies that $\Omega_{\Lambda}$ is determined by the choice of $\Omega_k$ and $\Omega_M$ through eq.~\ref{redsurface}. (ii) The observed density of luminous matter in the universe only amounts to $\Omega_M=0.005$, and the total baryonic matter density can be estimated to be at most $\Omega_M=0.05$. Any larger value of $\Omega_M$ implies the existence of dark matter.  (iii) Heuristic analysis cosmic microwave background power spectrum in the context of THED gravity (see next section) suggests that the universe might be negatively curved ($ k<0, \, \Omega_k > 0$); however in this analysis of fitting the supernovae data, all values of $\Omega_k$ shall be considered as equally admissible without prejudice.

Figure \ref{ThedFits} shows the  model-data discrepancy as a function of $\Omega_k$ for certain chosen values of $\Omega_M$. The green line corresponds to the intersection of the green and red surfaces in fig.~\ref{paramfigure}. The oscillating solutions lie on the left side of the green line and the ever-expanding solutions lie on the right side of the green line of fig.~\ref{ThedFits}. For any given $\Omega_M$ there exists $\Omega_k$ that minimizes the model-data discrepancy, but it lies on the left side of the green line wherein the universe bounces off from $a_{min}$ to expand forever. For any $\Omega_M <0.3$, the best fit values have $e \simeq 11$,  comparable to the best fit value from $\LCDM$. 
For larger $\Omega_M$ the best fit value of $e$ grows larger, however that is irrelevant given the observed matter density in the universe.

Figure \ref{ThedEr} illustrates the model accuracy across the bins for three values of $\Omega_M$ corresponding to  luminous matter ($\Omega_M =0.005$),  baryonic matter  ($\Omega_M =0.05$ ), and dark matter in $\LCDM$ ($\Omega_M = 0.25$).  All panels in the left column have a tiny negative $\Omega_{\Lambda}$, so that the parameters lie at the intersection of the green and red surfaces of fig.~\ref{paramfigure}. Among the oscillatory solutions, these are the solutions that best fit the data for any given $\Omega_M$. The smaller the choice of $|\Omega_{\Lambda} |$, the larger will be $a_{max}$ and $\tau$; however they cannot be determined by fitting the parameters to high precision with supernova data because no significant change occurs in the solutions for $a(t)$ at $t<0$ (past) for $|\Omega_{\Lambda} | \le 0.01$. So, based on the fits we could for instance place an upper bound on $|\Omega_{\Lambda}|$ and infer from fig.~\ref{Sizefig} that for a universe with  $\Omega_M = 0.25$, $a_{max}$ is at least 10 times the current size and the periodicity $\tau$ is at least 50 times the current Hubble time $\simeq$ 1 trillion years.       
 
The middle column in figure \ref{ThedEr} shows the best fit solutions that correspond to the various minimas in fig.~\ref{ThedFits}.   These solutions lie to the left of the green line in figure \ref{ThedFits} and are ever-expanding single bounce solutions. Their $e$ values are at par with that of the best fits from $\LCDM$ (compare with fig.~\ref{FRWer}). They all have $\Omega_{\Lambda} >0$, and $\Omega_k <0$ for larger $\Omega_M$.  
For a spatially flat universe the best fit solution has $ \{\Omega_M = 0.15,\,  \Omega_k =0, \, \Omega_{\Lambda} =  1.4, \, e \simeq 10.95 \}$, which is at par with best fit of $\LCDM$ $ \{\Omega_M = 0.27,\,  \Omega_k =0, \, \Omega_{\Lambda} =  0.73, \, e \simeq 10.95 \}$.

\section{Spatial curvature and CMB Data}
\label{CMB}

It is often claimed that the  cosmic microwave background (CMB)  provides an independent geometric confirmation that the universe is  spatially flat \cite{ooba2018planck}.  The power spectrum of the CMB radiation shows the first acoustic peak at  a multipole moment of $\ell \simeq 220$, and this feature is interpreted as a strong evidence for a flat universe with $\Omega_k \simeq 0$. This is however a very model-dependent interpretation from the $\LCDM$ perspective, which is not valid in the context of THED gravity. 

Two quantities need to be computed from the cosmological model in order to predict the position of the first acoustic peak of CMB power spectrum.  (i) The comoving size of the sound-horizon $h_d$ at the time of photon-baryon decoupling that occurs when the universe was at a temperature of 3000 Kelvin. This corresponds to a redshift $z_d \simeq 1100$ at a time $t_d$ when the size of the universe was  $a_d = 1/(1+z_d) \simeq 10^{-3}$.  (ii) The comoving angular diameter distance from decoupling surface, which is the radial coordinate $r_d$ from which the decoupled photons at time $t_d$ free-stream to reach us at the present epoch.

The sound horizon $h_d$ gives an estimate of the size of the physical universe that could have interacted through the acoustic waves emanated from primordial density fluctuations. Diffusion processes will smoothen out the density fluctuations at scales smaller than $h_d$, leaving behind the fluctuations of the scale of $h_d$ frozen in the CMB.
The initial density perturbations  should not be considered as point-sources, rather they are matter clumps of some characteristic size $\xi_d$ formed during the prior cycle of the universe ($\xi_d$ would be zero in the big bang scenario of \LCDM). When the universe contracts to the size $a_d$ in the prior cycle, the temperature would have reached 3000 Kelvin and the universe would have been ionized to a plasma state, at which point acoustic waves would be initiated. These waves would traverse through the universe as it contracts to a minima and re-expand back to size $a_d$, at which point the universe would again neutralize and the photons decouple to form the CMB.

Following the calculation in \cite{hu1997physics,hu2001cosmic}, we expect the angular scale of a typical hot-patch in the CMB sky would be $\theta_d = (h_d + \xi_d)/ r_d $, and the first peak to be at an angular moment $\ell = (1-\phi) \pi/ \theta_d$ \cite{doran2002location}. The shift parameter $\phi= 0.265$ in $\LCDM$  \cite{ page2003first}, and for simplicity we shall assume the same value in THED gravity.  

It is important to note the following geometric fact--Compared to a spatially flat universe  ($\Omega_k =0$), in a closed universe ($\Omega_k <0$) a hot-patch of a given size would appear larger with a power spectrum peak at a smaller $\ell$, while in an open universe ($\Omega_k>0$) the hot-patch would appear smaller with a peak at a larger $\ell$.  
 
Now the radial coordinate $r_d$ from which the decoupled photons at time $t_d$ would reach us at the present epoch can be computed from
 \begin{equation}
 \chi_d  = \int_0^{r_d} \frac{dr}{\sqrt{1+\Omega_k r^2}} = \int_{t_d}^{0}  \frac{ dt}{a(t)}  =  \int^{z_d}_{0} \frac{dz}{(\dot{a}/a)}  \label{rd}
 \end{equation}
 For a given set of parameters $\{ \Omega_M, \Omega_k, \Omega_{\Lambda} \} $, the value of $\chi_d$  calculated in THED gravity is smaller than the value calculated from the standard model. This can be seen by rewriting the term $(\dot{a}/a)$ in terms of redshift and availing the dynamical equations.   
\begin{eqnarray}
\chi_d  & \xrightarrow{\textrm{GR}} & \int^{z_d}_{0} \frac{dz}{\sqrt{\Omega_M(1+z)^3 + \Omega_k (1+z)^2  + \Omega_{\Lambda}}}  \nonumber \\ 
& \xrightarrow{\textrm{THED}} &  \int^{z_d}_{0} \frac{dz}{\sqrt{\Omega_M(1+z)^3 + \Omega_k (1+z)^2 + \Omega_{\Lambda} - \left[ \overset{..}{a}/a \right] } } \nonumber
\end{eqnarray}  
 Except in regions very close to minima, the value of $\left[ \overset{..}{a}/a \right] $ is negative. Hence $\chi_d$ calculated from the standard model can be treated as an upper bound for its value in THED.  

 Next to calculate the comoving horizon size $h_d$, note that the speed of sound in plasma $c_s =c/\sqrt{3 +9 \rho_b/4\rho_r}$, where $\rho_b, \rho_r$ are the baryon density and radiation density. When the universe is very small we can ignore the effect of curvature and dark energy, but we have to include the effect of radiation $\Omega_R$ ($\simeq 5 *10^{-5}$ of CMB photons). 
 \footnote{Neutrino contribution would increase $\Omega_R$ to $9 * 10^{-5}$ in \LCDM, but this may not be the case in THED gravity.}        
 \begin{eqnarray}
h_d & \xrightarrow{\textrm{GR}} & \int_{t_{o}}^{t_d} \frac{ c_s\, dt}{a(t)}  = \int_{z_d}^{\infty} \frac{c_s \,dz}{\sqrt{\Omega_M(1+z)^3 + \Omega_R (1+z)^4  }}  \nonumber \\
& \xrightarrow{\textrm{THED}} & 2 \int_{t_{min}}^{t_d} \frac{ c_s \,dt}{a(t)}  \qquad
\textrm{\{Contraction+Expansion\}} \nonumber
\end{eqnarray}
 Notice  that the horizon size computed in THED  will be significantly larger than the horizon size computed from \LCDM, primarily because the universe spends twice the time in the plasma state (considering the contraction phase and expansion phase around minima when $a< a_d$). With the inclusion of non-zero $\xi_d$, we should expect the size of a CMB hot-patch $(h_d +\xi_d)$ to be significantly larger than that expected from \LCDM.

 \begin{table}[]
\begin{tabular}{|c|c|c|c||c|c|}
\hline
&   $\Omega_M$ & $\Omega_k$  & $\Omega_{\Lambda}$ & $\ell_{peak} \,\{ \xi_d = 0\}$     \\
 \hline 
 
 a & 0.25  & ~ -0.65  & 2.3  & 37 \, \{70\}  \\
  
 b &  0.25  & ~ 0  & 1  &  107  \, \{199\}  \\
   
c &  0.25  & ~ 0.5  & 0  & 150   \, \{280\} \\
   
d &  0.25  & ~ 1.4  & -1.8  & 220   \, \{404\} \\ 
   
\hline 

 e &  0.15  & ~ 0  & 1.4  & 127  \, \{232\}  \\

 f & 0.15  & 0.7  & 0.0 & 228   \, \{416\} \\

\hline 

g &   0.10  & 0.4  &  0.8 &  230  \, \{420\}  \\

h &  0.10  & 0.8  &  0.0 &  307   \, \{557\} \\  

i&   0.10  & 1.5  &  -0.4  & 340  \, \{615\}  \\

\hline  

j &  0.05  &  0.1  &  1.6 & 227  \, \{407\} \\  
  
k &  0.05  &  0.65  &  0.5 & 415  \, \{640\} \\
   
l &  0.05  &  0.9  &  0.0 & 480  \, \{860\} \\

  \hline
\end{tabular}
\caption{Position of the first CMB peak is estimated under certain assumptions. 
$\Omega_k$ is fine-tuned around the values given in the table such that the minima $a_{min}$ reaches $a_d * 10^{-4} $ without compromising the numerical precision. We have fixed the baryon density (which is part of $\Omega_M)$ to be 0.05, the radiation density $\Omega_R$ to be $5*10^{-5}$ and $\xi_d$  to be 0.1 Mpc at the decoupling surface. For comparison, the peak position in the extreme case of $\xi_d=0$ is given in braces.
}
\label{CC}
\end{table}

Observations reveal that the first peak in CMB power spectrum lies at $\ell \simeq 220$ corresponding to an angular width $\theta_d$ of about 1 degree in the sky. This lays a tight constraint on $\LCDM$ model that $\Omega_k \simeq 0$. But in THED gravity, since  $h_d$ is significantly larger and $\xi_d$ is non-zero, the angular width of a typical hot patch $\theta_d = (h_d + \xi_d)/r_d$ can be accounted for only if $r_d$ is significantly larger. However, since the computed value of $\chi_d$ is smaller in THED, the only way $r_d$ can be made significantly larger is by making $\Omega_k$ significantly positive so that $r_d =  \sinh (\sqrt{\Omega_k} \chi_d) /\sqrt{\Omega_k} $.  Thus the very same constraint that demands the universe to be spatially flat in the $\LCDM$ model, essentially demands the universe to be spatially open in THED gravity. 
 
 An exact constraint on $\Omega_k$ cannot be derived without an assumption on the value of $\xi_d$. Naively, we may assume that the galaxies we observe in the universe at the present epoch would be dissociated from the Hubble flow because it is locked by self-gravity. So, during the previous contraction phase of the universe, we can expect the local group of  galaxies to have clumped together while roughly maintaining their size, and then dissolve into the plasma soup  when the universe shrinks to the size $a_d$. With this simplistic picture and considering the observed sizes of large galaxies observed today, we can conservatively expect $\xi_d$ to be at least 0.1 Mpc at the decoupling surface. Acknowledging that the size of galaxies vary widely and that we can only heuristically guess the value of $\xi_d$, the position of the peak of the CMB power spectrum $\ell_{peak}$ is computed for various suggestive parameter values in table \ref{CC}. 
 
 From table \ref{CC} it is clear that for any reasonable value of $\xi_d$, we require $\Omega_k$ to be significantly positive in order to account for the peak at $\ell \simeq 220$. In general,  larger values of $\Omega_{M}$ and larger values of $\Omega_{\Lambda}$ have an effect of reducing $\ell_{peak}$. Smaller values of $\Omega_M$ along with smaller (or negative) $\Omega_{\Lambda}$ would automatically fix $\Omega_k$ to be positive by eq.\ref{redsurface}. Notice that even with $\xi_d =0$, an $\Omega_M = 0.15 $ requires $\Omega_k =0$ to attain a reasonable $\ell_{peak}$ (table \ref{CC}-e); the trend in table \ref{CC} then indicates that for any larger  value of $\xi_d$, we would require a smaller $\Omega_M$ and a larger $\Omega_k$.
  
In principle, for larger $\Omega_M$ we can arrive at the appropriate $\ell_{peak}$ by making $\Omega_{\Lambda}$ significantly negative (table \ref{CC}-d), but this will make the universe oscillate very rapidly (not supported by the Supernovae data). On the other hand, we could also arrive at the appropriate $\ell_{peak}$ with smaller $\Omega_M$ values by compensating it with a significantly positive $\Omega_{\Lambda}$ (table \ref{CC}-j), however this would not correspond to oscillatory universe solutions. Nevertheless, note that in both cases  $\Omega_k$ is positive.  

For $\xi_d$ larger than our conservative estimate, $\ell_{peak}$ would be smaller than the values in table \ref{CC}, which will shift the preference to smaller values of $\Omega_M$. Since the calculation depends on the choice of $\xi_d$ and $a_{min}$, an exact numerical constraint cannot be placed on the value of $\Omega_k$. Nevertheless, based on the trend set in table \ref{CC}, we can conclude that $\Omega_k$ has to be significantly positive (negatively curved open universe). Taken along with the Supernovae data fits, we would require an $\Omega_M$ of the order of 0.05 to 0.15 and a preferably tiny negative dark energy to keep the universe oscillating.          

\section{Discussion}

The field equations of THED gravity (eq.~\ref{cosm}) allows for nonsingular cosmological solutions wherein the universe bounces from an arbitrarily minimum size without violating the Null Energy Condition (NEC). The energy densities of various components in the universe, namely matter $\Omega_M$, curvature $\Omega_k$ and dark energy $\Omega_{\Lambda}$, play a succinct role in distinguishing the various scenarios of whether the universe is ever-expanding from a bounce or if it is oscillating (section \ref{profile}). The minimum size of the universe $a_{min}$ depends on the fine balance between the energy densities and how close they lie to the red-surface of fig.~\ref{paramfigure}. At its minimum, the universe should be very hot (say at least 150 MeV--QCD phase transition temperature, requiring  $a_{min}  \simeq 10^{-12}$) so that nucleosynthesis could occur afresh every cycle; while at the same time it should be much larger than the Planck scale ($a_{min}  \gg 10^{-32}$) so as to avoid entering the quantum gravity regime. The maximum size of the universe and its periodicity in the oscillatory scenario explicitly depends on the tinyness of the negative dark energy (eq.~\ref{tiny}). With a heuristic numerical bound of $-0.01<\Omega_{\Lambda} < 0 $, we can expect the oscillation cycle to be at least 1 trillion years long. 
An intriguing feature of the oscillatory solutions is that irrespective of the size and periodicity of the oscillations, the average energy within each oscillation turns out to be precisely zero.

In the oscillatory universe scenario, the minimum size at which the universe bounces back from contraction phase to expansion phase  determines how hot the universe gets. The smaller and hotter the universe gets, the hot plasma soup could better dissolve the inhomogeneities aggregated from gravitational clumping in the prior cycle.  Black holes formed in prior cycles can be particularly resistant to such dissolution into plasma soup. THED gravity presents a plausible solution by preventing the extreme inhomogeneity-buildup into black holes. It has been shown that there can exist extremely compact objects satisfying the NEC, which would otherwise be expected to collapse into black holes in the context of GR \cite{Shankar2017}. That is, finite pressure can restrain the gravitational collapse of a massive object even after it has shrunk to an arbitrarily small size, thus preventing the formation of singularity. Hence in THED gravity it is possible to interpret the black holes as super-compact nonsingular objects that satisfy NEC, and such dense compact objects can be expected to dissolve into the plasma soup if the universe oscillates to a sufficiently small $a_{min}$, thus preventing the inhomogeneity-buildup over cycles.     

Among various alternate theories of gravity, THED gravity has the theoretical advantage of avoiding the  big bang singularity at the classical level without needing to resort to an unknown quantum gravity theory. 
Consequently, we do not have to resort to any inflationary scenario in the early universe to address the horizon problem, because the universe is infinitely old and the entire observable universe would have been in causal contact when we consider the times before the minima of the current cycle.  
These theoretical features however cannot be quantitatively evaluated for a reality-check due to their inaccessibly high energy scales. Practically, we can evaluate the model for its consistency with the expansion profile of the universe at late times (small redshift $z$) which is described by the Type Ia Supernova data.  
To discriminate between different models it can be very useful to describe the late times expansion profile of the universe in a  model-independent way as suggested by \emph{cosmographic analysis} wherein observables are expanded as  polynomial series in $z$ which can then be directly compared to the data \cite{capozziello2019extended}. Different models would yield different coefficients along the cosmographic series, thus discriminating one from the other. 
This technique is particularly useful to discriminate between action-modified models such as the $f(R)$ theories that attempt to feature the dark energy behavior without explicitly introducing the cosmological constant $\Lambda$ as a model parameter. 
In fact it is possible to phenomenologically construct an optimal functional form for $f(R)$ expanded as a power series in $z$ by matching it to the cosmographic series at each order \cite{capozziello2019extended} and connect the early and late epochs of the universe \cite{benetti2019connecting}. 
Analogously, it should in principle be possible to cosmographically construct more general $f(R)$ actions (without an explicit $\Lambda$ parameter) in THED framework  that is observationally consistent with data at small $z$ and also averts big bang at extremely large $z$.
However, here we only deal with the simple Einstein-Hilbert action ($f(R)$ is just $R$) in THED framework with $\Lambda$ as an explicit parameter for dark energy; hence we simply resort to fitting the model parameters to the Supernova data in a straightforward manner (section \ref{SUP}). 

The  binned analysis approach that we have adopted in section \ref{SUP} is simplistic but powerful enough to evaluate the model-data discrepancies at different scales independently, because the dataset we use is large enough to well-populate the bins at all scales.  The error measure $e$ (eq.~\ref{modelaccuracy}) depends on the binwidths which are chosen so that there are around 100 data points within each bin on average (see bottom right panel of  figure \ref{Datafig}), guided by an optimistic assumption that the uncertainties within each bin would wash away. More sophisticated evaluation criteria can be adopted to extract the best-fit parameters, but the binned analysis approach  used here is sufficient to reveal a birds-eye view of the data-fit accuracy over different  regions of the parameter space. As can be seen from fig.~\ref{ThedFits}, the best fit ever-expanding solutions require dark matter and positive dark energy, while the best fit oscillatory solutions require little to no dark matter with a small negative dark energy and a significant negative curvature (positive $\Omega_k$). 

The true model parameters should ideally be concordant with data from all the existing cosmological observations, not just the supernovae data. It is worth noting that when the supernovae data is considered in isolation, it is very consistent with the scenario of constant expansion rate  of the universe \cite{nielsen2016marginal}. However this scenario is not acceptable within the framework of $\LCDM$ because it requires the universe to be negatively curved, which turns out to be extremely unlikely when CMB data is also taken in account. But this scenario is feasible in THED framework because the CMB spectrum's first peak position indeed suggests a negative curvature (section \ref{CMB}). 

Recent improvements in analysis of CMB data released by PLANCK satellite in 2018 reveals with over 99 percent confidence level that the universe is positively curved with $-0.007 > \Omega_k > -0.095 $ and best fit at $\Omega_k =-0.04$ \cite{di2020planck}. Note that this detailed analysis has only marginally pushed the consensus value from $\Omega_k \sim 0$ to -0.04, but that is sufficient enough to reveal some serious discordances in the estimation of $\LCDM$ parameters \cite{di2020planck}. A similar detailed analysis could be performed in THED framework, which might marginally shift the value of $\Omega_k$ obtained from the preliminary analysis of the first peak position of CMB spectrum (section \ref{CMB}); however that would not  be drastic enough to affect the overall positivity of  $\Omega_k$.

Although an explicit numerical constraint cannot be placed on  $\Omega_k$ independently, fig.~\ref{ThedFits} shows that its positivity goes hand in hand with $\Omega_M$ being low (of the order of just the baryonic density) without requiring much dark matter.  This raises the question of whether THED gravity can account for phenomena like gravitational lensing and flat galactic rotation curves without requiring much dark matter. Interestingly, the gravitational field profile of a compact massive object in THED gravity \cite{Shankar2017} has certain notable features to contribute in this regard. At large distances the gravitational field resembles that of an equally massive black hole in GR, while at intermediate distances the gravitational field can be significantly larger than that of a black hole, and stay flat at intermediate to short distances. Hence the light rays passing close enough to the massive object can bend more than what we would expect in GR. It is conceivable that the gravitational lensing effect observed in the universe could potentially be explained in THED gravity with much less dark matter than that required in GR. Similarly, it is also conceivable that the flat galactic rotation curves can be accounted for  with much less dark matter in THED gravity than in GR if the galactic centers are modeled as such compact nonsingular objects. These speculations need further detailed analysis for their validation. Furthermore, whether structure formation in the universe can be achieved with less dark matter in THED gravity  needs to be analyzed.

\section*{Appendix}

 Consider the equation of motion (eq.~\ref{cosm}) around the minima where there is effectively only one dominant energy component (given by $n$).  
\begin{equation}
    \left[ \overset{..}{a}/a \right] +(\dot{a}/a)^{2}   =  \Omega /a^n .
    \label{A1}
\end{equation}   
Let the minimum occur at $t=0$ so that  $a(0) = a_{min}$. Defining  $\tilde{a} = a/a_{min}$ and $\tilde{t} =t \sqrt{\Omega / a_{min}^n} $. For $\tilde{a} \gg 1$, we can expect a power-law solution
\[ 
\tilde{a}  = \alpha \, \tilde{t}^{\beta} .
\] 
By substituting this solution back into the equation, we obtain
\begin{equation}
\beta (\beta -1) \left(  \frac{\tilde{t}^{\beta -2}}{\tilde{t}^{\beta}} \right)  + \beta^2 \left(  \frac{\tilde{t}^{\beta -1}}{\tilde{t}^{\beta}} \right)^2  = (\alpha \tilde{t}^{\beta})^{-n}
\end{equation}
\begin{equation}
 \Rightarrow (2 \beta^2  - \beta ) \tilde{t}^{-2} = \alpha^{-n} \tilde{t}^{-n \beta}  
\label{eqbeta}
\end{equation}
\begin{equation}
 \Rightarrow  n \beta =2, \qquad \alpha^{-n} = (2 \beta^2 - \beta ) . 
 \label{solbeta}
 \end{equation}

 Clearly this is not an exact solution, its only an asymptotic possibility. For all $n<4$, the solution indeed converges to the asymptotic limit. As $n \rightarrow 4$ it takes a very long time to reach the asymptotic limit, but for $n=3$ the convergence is rather quick. In the asymptotic limit,
\[ a/a_{min} \simeq \alpha \, t^{\beta} \, \Omega^{\beta/2} a_{min}^{-n \beta/2} \, , \]
the explicit value of $a_{min}$ factors out of the equation (because $n \beta =2$) and hence $a_{min}$ does not affect the solution as long as $a \gg a_{min}$  
\begin{equation} 
\Rightarrow a \simeq \alpha \, t^{\beta} \, \Omega^{\beta/2}  \qquad 
 t \simeq \frac{\alpha^{-n/2} a^{n/2} }{ \sqrt{\Omega} } .
\label{assymp}
 \end{equation}
Interestingly, this would be the exact solution for GR equations (without $\overset{..}{a}$ term in eq.~\ref{A1}) with $\alpha^{-n} = \beta^2$ and $n \beta =2$ instead of eq.~\ref{solbeta}. 
   
The solution is more complicated when $n>4$ (not a physically relevant situation). It turns out that for any $n>4$, the value of $\beta$ saturates at $1/2$ and the value of the $\alpha$ can only be numerically estimated. This is because eq.~\ref{eqbeta} cannot be solved to obtain $n \beta =2$ as in eq.~\ref{solbeta} if $\beta=1/2$. In this situation, the asymptotic convergence to the power-law solutions can be better analyzed by rewriting the equation  in terms of $ e^f \equiv \tilde{a}^2$ and $e^{\tau} \equiv \tilde{t}$, to obtain the following differential equation for $f(\tau)$.
\begin{equation}
[f'' + f'^2 -f'] = 2 e^{[2 \tau -nf(\tau)/2 ] }. 
\end{equation}
For convenience let's shift the time axis such that the minima occurs at $\tilde{t}=1$ rather than at 0, as we are interested only in the asymptotic limit when $\tilde{t}$ is very large. So the  initial conditions for the above equation are $f(\tau=0) =0$ and $f'(\tau=0)=0$. 
Numerically it turns out that  $f'' \rightarrow 0$ for large $\tau$. This implies that  $f(\tau) \rightarrow \gamma \, \tau + c$, and 
\begin{equation}
[\gamma^2 -\gamma] = 2 e^{[2 -n \gamma/2 ]\tau } \, e^{-cn/2}. 
\end{equation}
It can be readily seen that $\gamma$ cannot be less than 1. For all $n<4$, $\gamma = 4/n$ and $e^{-cn/2} =[\gamma^2-\gamma]$. While for $n>4$, we have $\gamma=1$ and the term $e^{[2 -n \gamma/2 ]\tau } \rightarrow 0$ for large $\tau$. The asymptotic solution is then 
\[ \tilde{a} =e^{(c/2)} \, \tilde{t}^{(\gamma/2)} . \] 
Thus for any $n>4$, $\gamma=1$ is equivalent to $\beta=1/2$ and the value $\alpha=e^{(c/2)}$ can be numerically computed to obtain the asymptotic solution.

\end{document}